\documentclass[12pt,reqno]{amsart}

\usepackage{amssymb}
\usepackage{amsthm}
\usepackage{amsmath}
\usepackage{dsfont}
\usepackage{fullpage}

\DeclareMathOperator*{\argmin}{argmin}

\usepackage{color}
\usepackage{graphicx}
\usepackage{wrapfig}
\usepackage{subcaption}
\usepackage{float}
\usepackage{algorithm2e}
\RestyleAlgo{ruled}

\renewcommand{\vec}[1]{#1}
\newcommand{\R}{\mathbb{R}}
\renewcommand{\P}[1]{\mathbb{P}\left(#1\right)}
\newcommand{\E}[1]{\mathbb{E}\left[#1\right]}
\newcommand{\Pb}{\mathbb{P}}
\newcommand{\Qb}{\mathbb{Q}}
\newcommand{\1}{\mathds{1}}
\newcommand{\dd}{\text{d}}

\theoremstyle{definition}

\theoremstyle{remark}

\numberwithin{equation}{section}

\begin{document}

\title{Universal approximation of credit portfolio losses using Restricted Boltzmann Machines}

\author{Giuseppe Genovese}
\address{Institute of Mathematics, Universit\"at Z\"urich, Switzerland}
\curraddr{}
\email{}

\author{Ashkan Nikeghbali}
\address{Institute of Mathematics, Universit\"at Z\"urich, Switzerland}
\curraddr{}
\email{}

\author{Nicola Serra}
\address{Department of Physics, Universit\"at Z\"urich, Switzerland}
\curraddr{}
\email{}

\author{Gabriele Visentin}
\address{Department of Mathematics, ETH Z\"urich, Switzerland}
\curraddr{}
\email{}


\date{February 20, 2022.}



\begin{abstract}
We introduce a new portfolio credit risk model based on Restricted Boltzmann Machines (RBMs), which are stochastic neural networks capable of universal approximation of loss distributions. 

We test the model on an empirical dataset of default probabilities of 1'012 US companies and we show that it outperforms commonly used parametric factor copula models -- such as the Gaussian or the t factor copula models -- across several credit risk management tasks. In particular, the model leads to better fits for the empirical loss distribution and more accurate risk measure estimations. 

We introduce an importance sampling procedure which allows risk measures to be estimated at high confidence levels in a computationally efficient way and which is a substantial improvement over the Monte Carlo techniques currently available for copula models. 

Furthermore, the statistical factors extracted by the model admit an interpretation in terms of the underlying portfolio sector structure and provide practitioners with quantitative tools for the management of concentration risk. 

Finally, we show how to use the model for stress testing by estimating stressed risk measures (e.g. stressed VaR) under various macroeconomic stress test scenarios, such as those specified by the FRB's Dodd-Frank Act stress test.
\end{abstract}

\maketitle

\section{Introduction}
\label{sec:introduction}

Portfolio credit risk models play a fundamental role within many financial institutions, in particular for the estimation of capital requirements and the pricing of basket credit derivatives. 

Despite the variety of portfolio models that have been proposed over the years (see \cite{gordy2000comparative} and \cite{embrechts2015quantitative} for an overview), in practice they all rely on the same \emph{conditional independence framework} \cite{gordy2000comparative, koyluoglu1998reconcilable, frey2001modelling}. More specifically, in all such models obligors are assumed to default independently given a set of underlying factors, which can be either latent (i.e. statistically estimated) or observable (e.g. macroeconomic indicators). Models of this kind are most generally known as mixture models. 

The rationale behind mixture models is that factors are expected to provide a low-dimensional parametrization of the complex dependence structure of the entire portfolio, which can then be exploited for dimensionality reduction and the simplification of otherwise computationally infeasible procedures. These latent factors can furthermore be interpreted by practitioners in terms of systematic and idiosyncratic risk, thus connecting credit risk practice with the neighboring fields of portfolio selection and asset pricing.

Apart from this common modelling assumption, each individual credit risk model relies on particular parametric specifications for the factor distribution and the obligors' default probabilities. These parametric choices are almost never motivated on empirical grounds and are often influenced by considerations of mathematical tractability or statistical expediency.

For example, in the case of CreditRisk+, factor distributions are chosen in such a way as to obtain a closed-form formula for the probability generating function of the total portfolio losses \cite{boston1997creditrisk+}\cite{bluhm2016introduction}, without addressing the arbitrariness of this distributional choice.

Similarly, in Moody's public-firm EDF\textsuperscript{\texttrademark} model the distance-to-default values are first computed using a classical Merton model methodology -- which is well-known to underestimate short-term credit spreads \cite{zhou2001term, hull2000valuing, hull2001valuing} -- and are only subsequently readjusted by calibrating the term structure on an historical database of default frequencies \cite{sun2012public}, in what amounts to a methodological patchwork.

A notorious example of over-reliance on parametric assumptions was the widespread use of the single-factor Gaussian copula model, which - under additional simplifying assumptions \cite{vasicek2002distribution} - can be used to derive closed-form expressions for the price of common basket credit derivatives. The single-factor Gaussian copula gained broad acceptance among practitioners and was adopted as the standardized method in the Basel II capital adequacy framework \cite{basel2004basel}. Perhaps more damagingly, it also became the \emph{de facto} standard model for the pricing of mortgage-backed asset securities, such as CDOs, in the early 2000s \cite{burtschell2005comparative}, up to the 2007-2008 financial crisis.

This over-reliance on parametric assumptions and lack of model validation in credit risk has been repeatedly shown to result in severe model misspecification across a range of financial applications. For instance, in CDO pricing it was pointed out very early that the single-factor Gaussian copula is unable to reproduce the observed market prices for tranches of different seniority, leading to the so called \emph{correlation smile} (see \cite{burtschell2005comparative} for an assessment of the problem for various parametric portfolio models). Model misspecification is particularly dangerous in credit risk management, in which the wrong choice of copula may lead to inaccurate risk measures and capital requirements \cite{gordy2000comparative, embrechts2015quantitative, frey2001copulas}.

A substantial amount of research has focused on addressing these concerns by identifying alternative parametric families that are believed to be better suited for credit risk applications. So, for instance, the single-factor Gaussian copula in \cite{li2000default} has been superseded by stochastic correlations models \cite{andersen2004extensions, schloegl2005modelling}, t copula models \cite{andersen2003all, demarta2005t, lindskog2000modelling, frey2003dependent, greenberg2004tuning, mashal2003inferring, schloegl2005note}, double t copula models \cite{hull2004valuation, cousin2008comparison}, Clayton copula models \cite{schonbucher2001copula, schonbucher2002taken, rogge2003modelling, madan2004credit, friend2005correlation}, Marshall–Olkin copula models \cite{duffie1998simulating, li2000default, wong2000copula, elouerkhaoui2003credit, giesecke2003simple, lindskog2003common}, more general Archimedean copulas \cite{nelsen2007introduction, genest1986joy}, and combinations of all the aforementioned models, via very general constructions in which pair copulas are glued together, as done in vine copulas \cite{bedford2001probabilistic, kurowicka2006uncertainty, aas2014bounds, czado2010pair}. 

Unsurprisingly, the question of how to choose the right copula has become a pressing practical problem for many practitioners and common recipes include trying to fit as many models as possible and to select the one with highest likelihood on any given dataset \cite{embrechts2015quantitative}.

In this paper we want to address these modelling questions by pursuing another approach and investigating instead the use of non-parametric, universal approximators from the field of unsupervised machine learning. These models are able to fit any dependence structure and therefore provide a way of mitigating the intractable - and ultimately data-dependent - issue of choosing the right parametric copula, by letting the data speak for itself. Crucially these models maintain appealing mathematical features of traditional credit risk model -- namely their conditional independence structure -- which can be exploited in numerous applications and substantially improve interpretability, as we will see.

While our approach can be successfully applied to many stochastic neural networks, such as Variational Autoencoders (VAEs) and Generative Adversarial Networks (GANs), in this paper we focus on a single model, the Restricted Boltzmann Machines (RBMs), in order to showcase as many applications to credit risk as possible. 

Section \ref{sec:introduction_to_RBMs} is a self-contained introduction to RBMs, with particular emphasis on their conditional independence structure and issues pertaining to the training of the model. 

In Section \ref{sec:estimation_of_risk_measures} we introduce the credit RBM model and we see how its universal approximation property allows us to fit an empirical dataset of default probabilities in a non-parametric way, with substantial improvements over common copula models. 

Section \ref{sec:importance_sampling} is devoted to the fundamental issue of importance sampling for credit risk models, which is pivotal for the accurate estimation of risk measures. We exploit the fact that the distribution of the credit RBM model admits an explicit formula and develop an efficient importance sampling procedure for the estimation of large losses. This result represents a substantial improvement over current standards, since no importance sampling procedure is known for portfolios other than the Gaussian copula model.

In Section \ref{sec:sector_structure} we deal with the issue of the interpretability of the latent factors extracted from data by the RBM model. In particular we show how the receptive field of the hidden units encodes information on the sector structure of the credit portfolio. This information can be represented graphically and be used by portfolio managers to monitor portfolio diversification and concentration risk.

Finally, in Section \ref{sec:stress_testing}, we see how the RBM model can be used to perform stress tests on credit portfolios, by training on a joint dataset of default probabilities and macroeconomic factors. The impact of stress test scenarios on the total loss distribution can then be assessed by performing conditional blocked Gibbs sampling. As an illustrative example, we implement the Federal Reserve Board's (FRB) Dodd-Frank Act stress test of June 2020 on a real portfolio. More generally, this methodology allows a portfolio manager to quantify the impact of any macroeconomic scenarios on portfolio performanceand to compute stressed risk measures (e.g. stressed VaR) for use in executive settings.

\section{Introduction to RBMs}
\label{sec:introduction_to_RBMs}

\subsection{Model description}

Restricted Boltzmann Machines (RBMs) are undirected probabilistic graphical models used for learning probability distributions. They are an instance of energy based models introduced in the 1980s \cite{ackley1985learning, hinton1983analyzing, hinton1986parallel, smolensky1986information} which can be trained efficiently \cite{hinton2002training} and played a fundamental role in the early stages of the deep learning revolution \cite{hinton2007learning, hinton2006reducing}. What follows is a short introduction to the model, but we refer the interested reader to \cite{hinton2012practical, bengio2013representation, fischer2014training, montufar2016restricted} for more detailed treatments.

\begin{figure}[ht]
    \centering
    \includegraphics[width=0.5\textwidth]{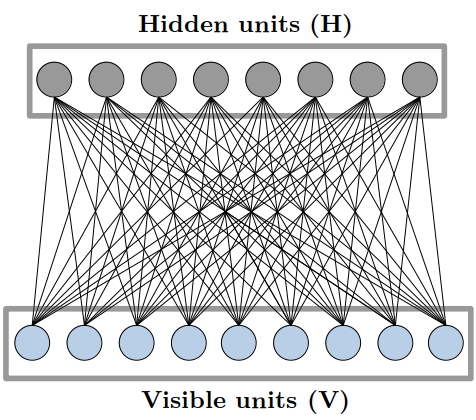}
    \caption{Graphical representation of a Restricted Boltzmann Machine.}
    \label{fig:RBM}
\end{figure}

The units of an RBM are partitioned into two layers: the layer of visible units, $\vec{V} = (V_1, V_2, \ldots, V_n)$, encoding the distribution of the training data, and the layer of ancillary hidden units, $\vec{H} = (H_1, H_2, \ldots, H_m)$. The two layers are fully connected, while intra-layer connections are absent, thus leading to a bipartite network, as depicted in Figure \ref{fig:RBM}. We associate to a RBM the following probability density:
\begin{equation}
\label{eq:distribution}
p(\vec{v}, \vec{h}; \theta) = \frac{1}{Z} e^{-E(\vec{v}, \vec{h};\theta)} p_V(v)p_H(h), \quad \vec{v} \in \R^n, \vec{h} \in \R^m\,,
\end{equation}
where $p_V$ and $p_H$ denote the a priori distributions of the units belonging respectively to the visible and hidden layer and
\begin{equation}
\label{eq:energy}
E(\vec{v}, \vec{h};\theta) = - \vec{h}^T W \vec{v} - \vec{b}^T \vec{v} - \vec{c}^T \vec{h}
\end{equation}
is the so-called \emph{energy function}. It is parametrized by a matrix of real-valued weights, $W = (W_{ji}) \in \R^{m \times n}$, and two bias vectors, $\vec{b}$ and $\vec{c}$ (for the visible and hidden units respectively), and we set $\theta=(W,b,c)$. The constant $Z$ is simply a normalization constant called the \emph{partition function} of the model. 

One important remark is that the bipartite structure of an RBM implies a corresponding factorization of the distribution function, so that an RBM can be seen as a mixture model, in which visible units are conditionally independent given the values of all the hidden units, and viceversa. To see this, we specialize to the case of uniform a priori distributions and binary units, i.e. $\vec{v} \in \{0,1\}^n$ and $\vec{h} \in \{0, 1\}^m$, as we will do for the rest of the paper, and compute the following conditional distributions:
\begin{align}
    \P{V_i = 1|H} & = \sigma \left( \sum_{j=1}^m W_{ji} H_j + b_i \right), \quad \forall i = 1, \ldots, n \label{eq:pds_in_RBM} \\
    \P{H_j = 1|V} & = \sigma \left( \sum_{i=1}^n W_{ji} V_i + c_j \right), \quad \forall j = 1, \ldots, m \label{eq:hidden_given_visible}
\end{align}
where $\sigma(\cdot)$ denotes the sigmoid function. 

We stress that the availability of an explicit formula for the joint distribution of the RBM's units is an important advantage over other generative models, such as Variational Autoencoders (VAEs) or generators trained via adversarial learning, as in Generative Adversarial Networks (GANs). In Section \ref{sec:importance_sampling} and Section \ref{sec:stress_testing}, we will see how to exploit this fact in the context of credit risk. 

Nevertheless, sampling from the RBM distribution (\ref{eq:distribution}) can be computationally inefficient. Indeed an exact computation of the normalisation constant $Z$ requires a number of operations that is exponential in the number of units, which makes exact sampling from the model distribution infeasible. One is thus led to the use of approximate methods, such as Markov chain Monte Carlo techniques.

Indeed Equations \eqref{eq:pds_in_RBM} and \eqref{eq:hidden_given_visible} naturally suggest the blocked Gibbs sampling procedure described in Algorithm \ref{alg:Gibbs_sampling}. 

\vspace*{0.5cm}
\begin{algorithm}[H]
\caption{Blocked Gibbs sampling for RBM.}
\label{alg:Gibbs_sampling}
\KwData{RBM with parameters $(W,\vec{b},\vec{c})$; $T=$ number of Gibbs steps.}
\KwResult{Sample $(\vec{v}, \vec{h})$ from RBM distribution.}
Sample $\vec{v}$ from the uniform distribution on $\{0,1\}^n$.\

\For{$k = 1$ \KwTo T}{
        Sample $\vec{h}$ given $\vec{v}$ using Equation \eqref{eq:hidden_given_visible}.\
        
        Sample $\vec{v}$ given $\vec{h}$ using Equation \eqref{eq:pds_in_RBM}.
}
\Return{$(\vec{v}, \vec{h})$.}
\end{algorithm}
\vspace*{0.5cm}

This procedure is very efficient, since all units in the same layer can be updated simultaneously. This is in fact the principal advantage of RBMs when compared to other energy models with more complex graph topologies.

\subsection{Training}

RBMs can be trained using likelihood maximization via stochastic gradient ascent. The gradient of the log-likelihood function with respect to any model parameter $\theta = (W, \vec{b}, \vec{c})$ satisfies the following general expression:

\begin{equation}
\label{eq:loglikelihood_gradient}
    \frac{\partial \mathcal{L}(\theta|\vec{v})}{\partial \theta} = \mathbb{E}_{\vec{H} | \vec{V}}\left[ \frac{\partial E(\vec{v}, \vec{h})}{\partial \theta} \right] - \mathbb{E}_{\vec{V}, \vec{H}}\left[ \frac{\partial E(\vec{v}, \vec{h})}{\partial \theta} \right]
\end{equation}

The first expectation, corresponding to the conditional distribution of $\vec{H}$ given $\vec{V}$, can be evaluated directly on the dataset (or on a mini-batch of training samples). The second summand in (\ref{eq:loglikelihood_gradient}) involves the distribution (\ref{eq:distribution}) and therefore it is hard to compute. We have to approximate this second term by a Monte Carlo method such as the blocked Gibbs sampling in Algorithm \ref{alg:Gibbs_sampling}. But unfortunately this routine would have to be used at each paraemeter update, which makes overall training computationally demanding, as full convergence to the model distribution may require in practice as many as $10^3$ Gibbs steps \cite{kondratyev2019market}.

The breakthrough by Hinton \cite{hinton2002training} was to observe that the model can be efficiently trained by evaluating the second expectation in \eqref{eq:loglikelihood_gradient} by running Algorithm \ref{alg:Gibbs_sampling} for only few Gibbs steps at each parameter update and this training method came to be known as Contrastive Divergence (CD). In this paper we employ a popular improvement of the CD algorithm, namely Persistent Contrastive Divergence (PCD) \cite{younes1999convergence, tieleman2008training}. PCD is reported schematically below, where for simplicity we present the version in which all the dataset is explored at each step. 

\vspace*{0.5cm}
\begin{algorithm}[H]
\caption{$k$-steps Persistent Contrastive Divergence.}
\label{alg:PCD}
\KwData{Initial parameters $\theta_0=(W_0,\vec{b_0},\vec{c_0})$; M datapoints $x_1,\ldots,x_M\in \R^n$; $T =$ number of iterations.}
\KwResult{Maximal likelihood estimator $\theta_T$}

Set $x_1^{[0]}=x_1\ldots x_M^{[0]}=x_M$\

\For{$\ell = 1$ \KwTo T}{
	\For{$j = 1$ \KwTo M}{
		For $p(\vec{v}, \vec{h}; \theta_{\ell-1})$ run Algorithm \ref{alg:Gibbs_sampling} for $k$ steps starting at $x_{[j]}^{[\ell-1]}$.\ 
		Store the final sample $x_{[j]}^{[\ell]}$.
		}
		Compute the approximated log-likelihood gradient
		$$
		\frac1M\sum_{j=1}^M\left(\mathbb{E}_{\vec{H} | \vec{V=x_j}}\left[ \frac{\partial E(\vec{v}, \vec{h})}{\partial \theta} \right] - \mathbb{E}_{\vec{H} | \vec{V=x^{[\ell]}_j}}\left[ \frac{\partial E(\vec{v}, \vec{h})}{\partial \theta} \right]\right)\,.
		$$
		Update $\theta_{\ell}\gets \theta_{\ell-1}$.
		
}
\end{algorithm}
\vspace*{0.5cm}



\section{The credit RBM model}
\label{sec:estimation_of_risk_measures}

In this section we introduce a new RBM-based static credit portfolio model, which we call the \textit{credit RBM model}. More specifically, we work with an RBM with binary units, in which the visible units are the default indicators of obligors in the portfolio (i.e. $V_i = \1_{\mathcal{D}_i}$ where $\mathcal{D}_i$ is the event that the $i$-th obligor defaults), so that the total number of defaults in the portfolio is given by $L_n = \sum_{i=1}^n V_i$. The hidden units instead are binary statistical latent factors, analogously to the factors in a mixture model.

In practice this model relies on a specific parametrization for the obligors' default probabilities, as is clear from Eq. \eqref{eq:pds_in_RBM}. Nevertheless, RBMs have been shown to be universal approximators for distributions on $\{0,1\}^n$ \cite{le2008representational, freund1994unsupervised, montufar2011refinements}, which implies that the credit RBM model is capable of learning loss distributions with any dependence structure, which is to say any copula. This flexibility allows us to dispense with parametric assumptions and can be shown empirically to yield better fits for portfolio loss distributions.

To test this claim we train a credit RBM model on an empirical dataset of daily one-year default probabilities (PDs) for 1'012 publicly listed US companies from 4 January 1999 to 30 December 2022. In order to check how our results depend on the credit rating of the obligors, we have partitioned the dataset into six portfolios (with size ranging from 40 to 300) of varying credit quality as measured using Moody's rating classes (Aaa, Aa, A, Baa, Ba, and B). A full description of the dataset is in Appendix \ref{sec:empirical_data}, see also Table \ref{tab:rating_classes} for a quick overview of the dataset breakdown by rating class. The model is trained independently on each portfolio. Whenever the results vary sensibly across rating classes, we mention it explicitly in the main text.

Our implementation of the credit RBM model is written in Python using, among others, the PyTorch package and runs on GPUs. The code is freely available on GitHub\footnote{Link: \texttt{https://github.com/gvisen/credit-RBM-model}. The dataset used for this paper comes from Bloomberg's Corporate Default Risk Model (DRSK) and is proprietary, therefore it could not be shared. But a toy dataset of default probabilities for 30 US listed companies computed by the authors using the Merton methodology is freely available on the GitHub repository, together with a training demo.}.

In order to train the binary RBM on a dataset of default probabilities, we replace the visible units by their activation probabilities (\ref{eq:pds_in_RBM}), as in \cite{hinton2006fast}. The RBM is then trained on the training dataset using Persistent Contrastive Divergence (PCD) for 100 Gibbs steps, 5000 epochs, a linearly decreasing learning rate with initial value $\eta = 2 \cdot 10^{-3}$, and mini-batch size equal to 250.

The number of hidden units is here chosen by $k$-fold cross-validation with $k=5$ and varies across portfolios, mainly depending on the size of the portfolio (Aaa: 250, Aa: 250, A: 1000, Baa: 1000, Ba: 500, B: 250). Model selection for generative models is more complicated than in the case of supervised learning models, because probability distributions admit several orthogonal similarity metrics \cite{theis2015note, wu2016quantitative, chen2018metrics, borji2019pros}. In our case, we have cross-validated the number of hidden units using the following metrics for model selection: KDE-based log-likelihood estimation \cite{goodfellow2014generative, theis2015note}, Maximum Mean Discrepancy (MMD) \cite{gretton2012kernel}, $L^2$ reconstruction error \cite{xiang2017effects}, and Number of statistically Different Bins (NDBs) \cite{richardson2018gans}.

\begin{figure}[h!]
\centering
\begin{subfigure}{.45\textwidth}
  \includegraphics[width=\linewidth]{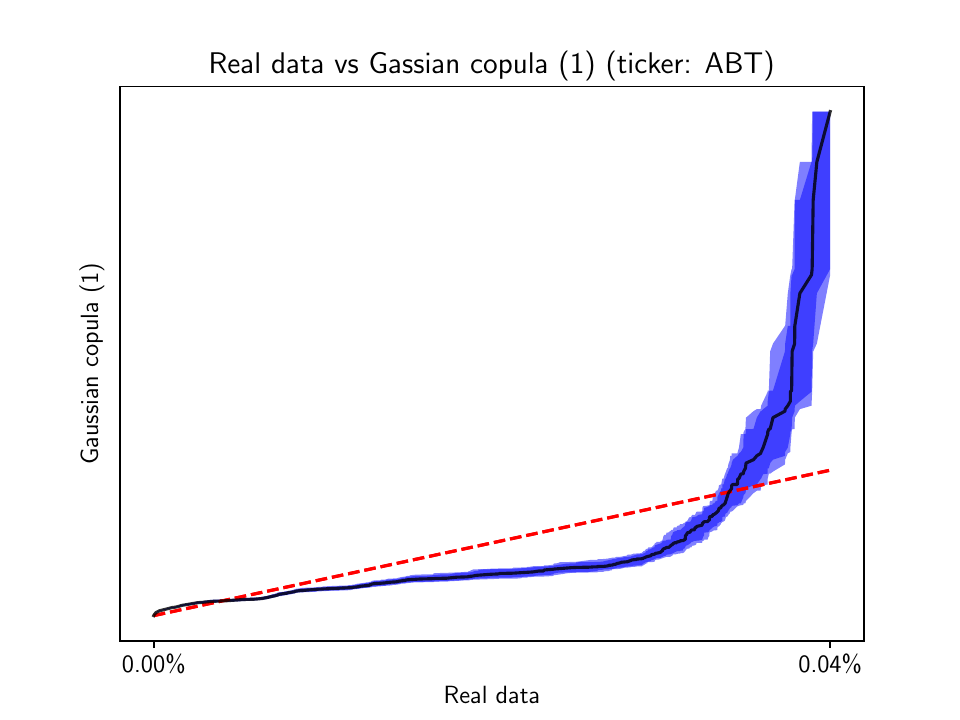}
\end{subfigure}
\begin{subfigure}{.45\textwidth}
  \includegraphics[width=\linewidth]{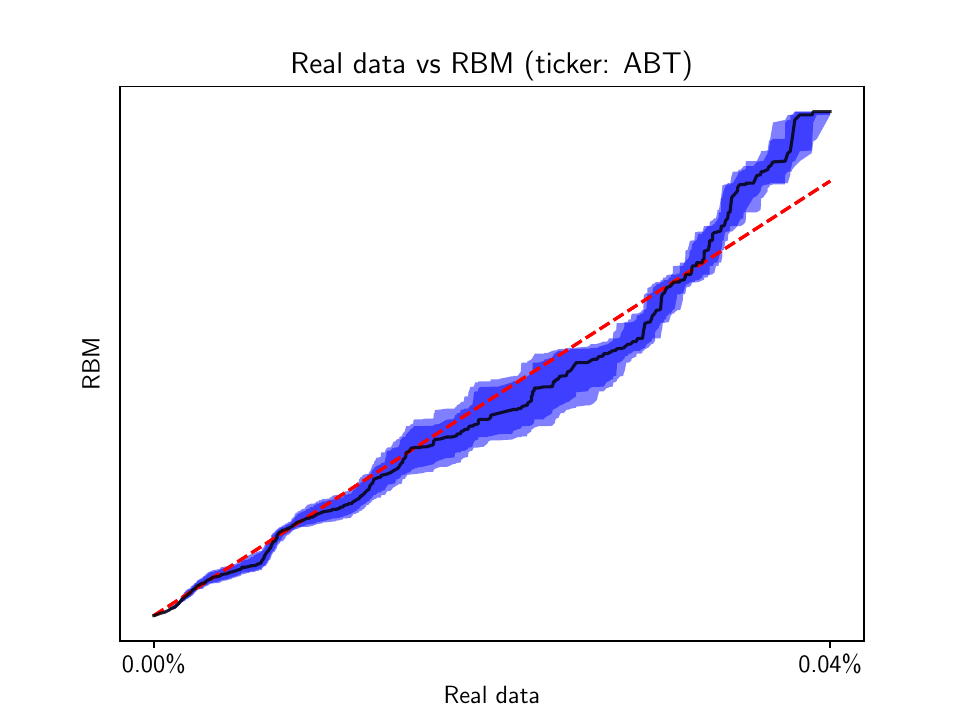}
\end{subfigure} 
\caption{QQ-plots of real and estimated default probabilities for Abbott Laboratories, using a single-factor Gaussian copula model (left) and the credit RBM model (right). The QQ-plots are shown within their 99\% and 95\% non-parametric bootstrap confidence bands (100 bootstrap samples).}
\label{fig:qq_plots_ABT}
\end{figure}

\begin{figure}[h!]
\centering
\begin{subfigure}{.45\textwidth}
  \includegraphics[width=\linewidth]{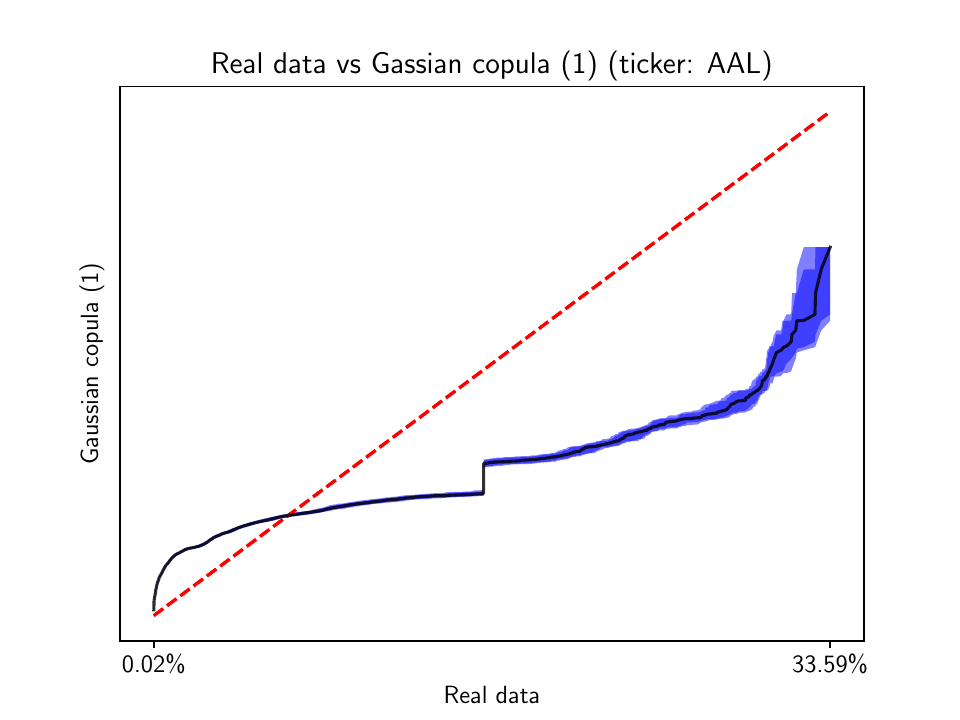}
\end{subfigure}
\begin{subfigure}{.45\textwidth}
  \includegraphics[width=\linewidth]{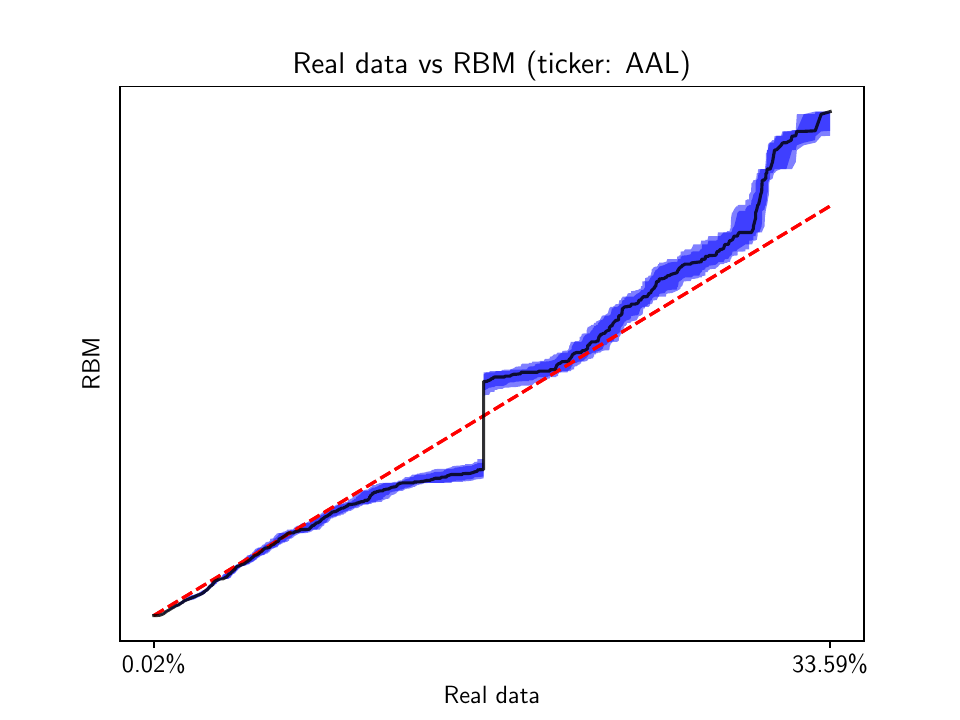}
\end{subfigure} %
\caption{QQ-plots of real and estimated default probabilities for American Airlines Group, using a single-factor Gaussian copula model (left) and the credit RBM model (right). The QQ-plots are shown within their 99\% and 95\% non-parametric bootstrap confidence bands (100 bootstrap samples).}
\label{fig:qq_plots_AAL}
\end{figure}

First of all, we focus on the ability of the RBM model to reproduce the marginal distributions of single-firm default probabilities.

Figures \ref{fig:qq_plots_ABT} and \ref{fig:qq_plots_AAL} show the QQ-plots comparing the distribution of real default probabilities with that generated by a single-factor Gaussian copula model (left) and the credit RBM model (right) for two representative companies in the dataset, Abbott Laboratories (ticker: ABT, rating: Aaa) and American Airlines Group (ticker: AAL, rating: B).

The single-factor Gaussian copula model with heterogeneous factor loadings has been fitted using the procedure presented in Appendix \ref{sec:fitting_copulas}. 

The marginal distributions are satisfactorily learned by the credit RBM model, while the single-factor Gaussian copula model fails to fit well the upper tail of most marginal distributions, as shown in the case of these two representative companies. 

The difference in performance is even starker if we look at the distribution of the aggregate portfolio losses. This distribution is statistically more difficult to learn, since portfolio losses depend on the full dependence structure among obligors' defaults.

\begin{figure}[H]
    \centering
    \includegraphics[width=\textwidth]{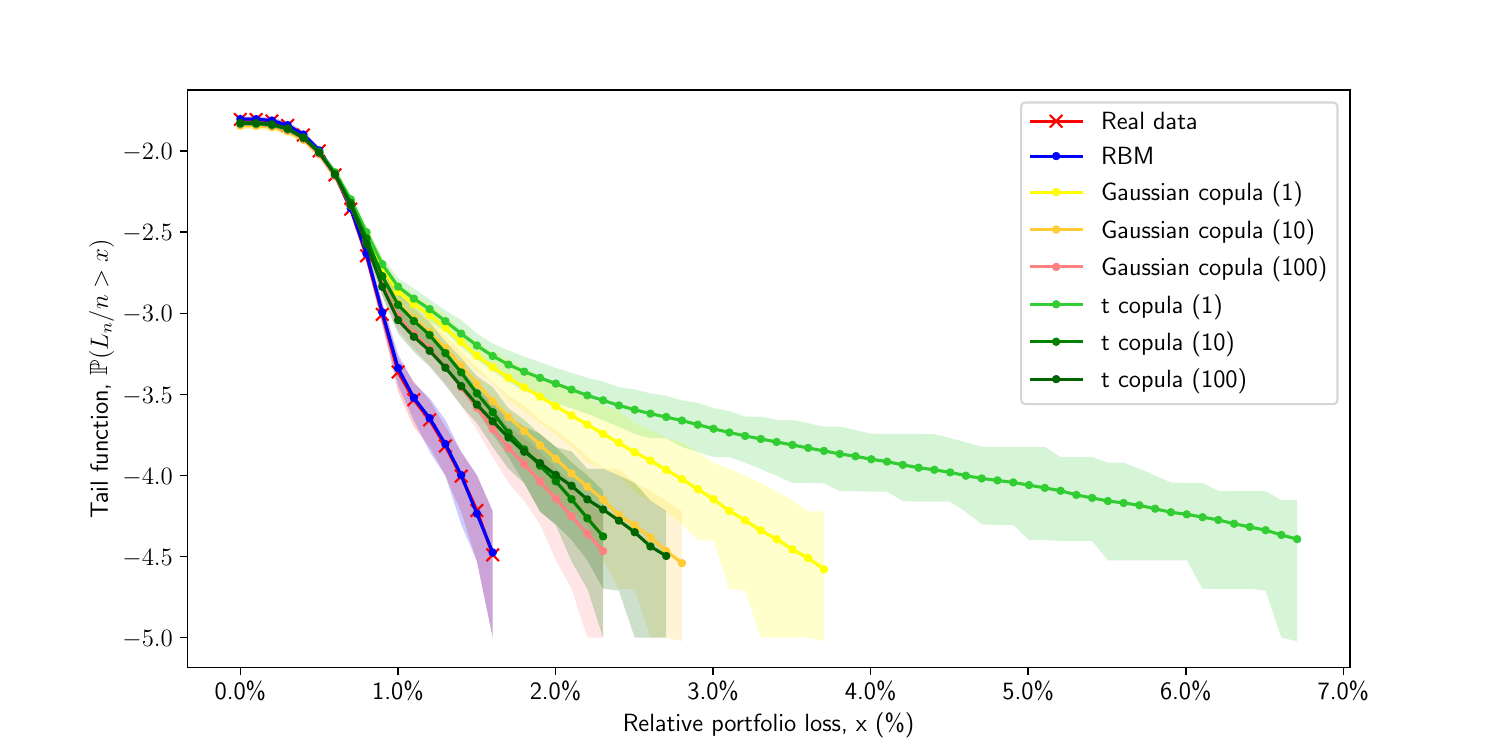}
    \caption{Comparison of tail functions of relative portfolio losses (rating Aa). Shaded areas correspond to 95\% non-parametric bootstrap confidence bands (100'000 Monte Carlo samples, 100 bootstrap samples).}
    \label{fig:tail_comparison_copulas}
\end{figure}

Figure \ref{fig:tail_comparison_copulas} shows a comparison of the tail probability function of relative portfolio losses as estimated from empirical data and from a selection of fitted models. 

Since we work with default probabilities, the tail function of losses needs to be estimated using Monte Carlo simulations. We further assumed that recovery rates are iid $\text{Beta}(4, 4)$ random variables. This sampling procedure introduces noise in the estimation of the tail function, which we quantify using a 95\% non-parametric bootstrap confidence band (shaded areas in the figures). 

We compare the credit RBM model, the Gaussian copula model, and the t copula model. We fitted copula models with a varying number of factors, in order to check whether more factors lead to an improvement in performance. We mention that in financial practice the number of factors used in factor copula models is typically very low, most commonly just one.

As the figure shows, the copula models overestimate the higher quantiles of the loss distribution, which translates into inaccurate portfolio risk measures, especially at high confidence levels. This overestimation is typically reduced when a richer factor structure is imposed, but persists even in models with as many as 100 factors. 

The poor performance of factor copula models is to be attributed to their parametric misspecification. The assumption of a multi-dimensional Gaussian or t-distributed latent asset process is unwarranted and it amounts to an arbitrary extrapolation of linear correlations (and, for the t-copula, also the degrees of freedom) to the entire dependence structure.

The credit RBM model, on the other hand, is able to match both the body and the tail of the empirical loss distribution remarkably well. The increase in performance is due to the universal approximation property of the RBM model, which makes it resilient to model misspecification.

We point out that Figure \ref{fig:tail_comparison_copulas} shows the results only for one of the six credit portfolios in our dataset, specifically the portfolio with Aa rating. Qualitatively similar results apply to all lower ratings (see Section \ref{subsec:tail_comparison_all_rating_classes}), but for the highest rating class, Aaa, we remark that the gap in performance between the credit RBM model and the factor copula model is strongly reduced. This indicates that the parametric assumptions underlying the Gaussian and t copula models may be adequate in the special case of highly-rated corporate portfolios.

\section{Importance Sampling for RBMs}
\label{sec:importance_sampling}

\subsection{Introduction to importance sampling in credit risk}

In this section we will see how a credit RBM model can be used to perform importance sampling (IS) estimation of the full tail function of portfolio losses in a computationally efficient way.

It is well-known that, even in the case of the simplest parametric credit risk model, such as the one-factor Gaussian copula model with equicorrelation structure, it is not possible to derive a non-asymptotic closed-form formula for the portfolio loss distribution. 
Semi-analytical or advanced simulation-based techniques are the only solution for estimating risk measures or tail functions at very high confidence levels.

Examples of semi-analytical methods are the saddle-point approximation \cite{martin2001saddle, gordy2002saddlepoint}, the Panjer recursion \cite{hull2004valuation, brasch2004note}, the Stein-Chen's method for Gaussian and Poisson approximation \cite{el2009stein, el2008gauss}, the large deviations principle \cite{dembo2004large} and, more recently, mod-$\phi$ approximation schemes \cite{meliot2022mod}.

As far as simulation-based methods are concerned, the starting point is the observation that a naive Monte Carlo approach is computationally expensive and is likely to underestimate risk measures, especially for high confidence levels. To overcome this problem, many authors have investigated the use of importance sampling techniques for portfolio credit risk models \cite{arvanitis2004credit, glasserman2004monte, glasserman2005importance, kalkbrener2004sensible, merino2004applying, egloff2005optimal}. The approaches developed so far apply exclusively to the Gaussian copula model and even in the Gaussian case they rely on approximate optimization, so that an extension to more general factor models appears difficult.

In the case of credit RBM models, instead, it is possible to perform efficient and accurate importance sampling estimation of the entire tail function of portfolio losses and it is therefore possible to compute risk measures at arbitrarily high confidence intervals. 

\subsection{Importance sampling for RBMs}

In order to perform importance sampling for a credit RBM model, we need to find a change of measure from the distribution of the visible units
$$
\Pb= {\mathbb E}_{H\,|\,V}[p(V,H;\theta)]
$$
to a new distribution, $\Qb$, under which losses are more frequent. It is natural to consider the change of measure suggested by the large deviations upper bound for $L_n = \sum_{i=1}^n V_i$, which leads to the following parametrized family $(\Qb_t)_{t \in \R^+}$:

$$ \frac{\dd \Qb_t}{\dd \Pb}(\vec{v}) = \frac{e^{t \sum_{i=1}^n v_i}}{\Gamma(t)}, \quad t \in \R^+$$

where $\Gamma(t)=\E{e^{t \sum_{i=1}^n V_i}}$ denotes the moment generating function of $L_n$. This is readily computed from Eq. \eqref{eq:distribution} and \eqref{eq:energy} as:

\begin{align}
\Gamma(t) & = \E{e^{tL_n}} \nonumber \\
& = \sum_{\vec{v} \in \{0,1\}^n} \sum_{\vec{h} \in \{0,1\}^m} \frac{1}{Z} \exp{\left( t \sum_{i=1}^n v_i \right)} \exp{\left( -E(\vec{v}, \vec{h}) \right)} \nonumber \\
& = \sum_{\vec{v} \in \{0,1\}^n} \sum_{\vec{h} \in \{0,1\}^m} \frac{1}{Z} \exp{\left( \vec{h}^T W \vec{v} + (\vec{b} + t\vec{u})^T \vec{v} + \vec{c}^T \vec{h} \right)} \nonumber \\
& = \frac{\tilde{Z}_t}{Z}, \label{eq:RBM_mgf}
\end{align}
 
where $\vec{u}\in\R^n$ is such that $u_i=1$ for all $i=1\ldots n$ and $\tilde{Z}_t$ denotes the partition function of an RBM with parameters $(W, \vec{b} + t\vec{u}, \vec{c})$.

To understand better the behavior of the tilted measure, $\Qb_t$, we further notice that:

\begin{align*}
    \frac{\dd \Qb_t}{\dd \Pb}(\vec{v}) & = \frac{Z}{\tilde{Z}_t} \exp{\left( t \sum_{i=1}^n v_i \right)} \\
    & = \frac{Z}{\tilde{Z}_t} \exp{\left( t \sum_{i=1}^n v_i \right)} \frac{\sum_{\vec{h} \in \{0,1\}^m} \exp{\left( -E(\vec{v}, \vec{h}) \right)}}{\sum_{\vec{h} \in \{0,1\}^m} \exp{\left( -E(\vec{v}, \vec{h}) \right)}} \\
    & = \left(\frac{1}{\tilde{Z}_t} \sum_{\vec{h} \in \{0,1\}^m} \exp{\left(t \sum_{i=1}^n v_i -E(\vec{v}, \vec{h}) \right)} \right) \bigg/ \left( \frac{1}{Z} \sum_{\vec{h} \in \{0,1\}^m} \exp{\left( -E(\vec{v}, \vec{h}) \right)} \right)
\end{align*}

Therefore under the measure $\Qb_t$ the visible units, $\vec{V} = (V_1, \ldots, V_n)$, follow the distribution of an RBM with parameters $(W, \vec{b} + t\vec{u}, \vec{c})$ and partition function $\tilde{Z}_t$. 

This exponentially tilted RBM is identical to the original RBM in all but the visible bias parameter, which is shifted to the right by $t$ and therefore results in higher portfolio losses, as can be deduced from Eq. \eqref{eq:pds_in_RBM} by recalling that the sigmoid function is strictly increasing.

If we are interested in estimating tail probabilities of the form $\P{\sum_{i=1}^n V_i > x}$ we can find the optimal tilting parameter, $t^*$, by minimizing the corresponding large deviations upper bound:

\begin{equation}
    t^* = \inf_{t \in \R^+} \left\{ \log{\Gamma(t)} - tx \right\} \:\: \Longrightarrow  \:\: t^* \:\: \text{such that} \:\: \mathbb{E}_{\Qb_{t^*}}\left[\sum_{i=1}^n V_i\right] = x
\end{equation}

In practice it is not necessary to tilt the RBM for each value of $x$ at each we want to evaluate the tail function. Rather, the entire probability tail function can be computed efficiently from a single tilted RBM (provided the tilting parameter, $t^*$, is sufficiently high), using the following IS estimator:

\begin{equation}
\label{eq:IS_estimator}
    \P{\sum_{i=1}^n V_i > x} \approx \frac{1}{M} \sum_{\ell=1}^M \1_{\{\sum_{i=1}^n v_i^{(\ell)} > x\}} \exp{\left(- t^* \sum_{i=1}^n v_i^{(\ell)}\right)} \frac{\tilde{Z}_{t^*}}{Z}
\end{equation}

where $(\vec{v}^{(\ell)} = (v_1^{(\ell)}, \ldots, v_n^{(\ell)}))_{\ell = 1}^M$ is a sample of visible units obtained from a tilted RBM with parameters $(W, \vec{b} + t^*\vec{u}, \vec{c})$.

\subsection{Estimation of ratios of partition functions}

One last numerical issue pertains to the computation of the ratio $\tilde{Z}_{t^*}/Z$ in Eq. \eqref{eq:IS_estimator}, which in general cannot be done exactly (with the exception of very small portfolios) due to the computational intractability of the partition function.

An efficient estimation procedure relying on Annealed Importance Sampling (AIS) was presented in \cite{salakhutdinov2008quantitative} and can be used to obtain fast and accurate estimations of this ratio. 

To briefly introduce this procedure, let us start by considering the simple Monte Carlo estimator for this ratio obtained from Equation \eqref{eq:RBM_mgf}, which is as follows:

\begin{equation}
\label{eq:partitions_ratio}
Z_{t^*}/Z \approx \frac{1}{M} \sum_{\ell = 1}^M \exp \left( t^* \sum_{i=1}^M v_i^{(\ell)} \right),  
\end{equation}

where $(\vec{v}^{(\ell)} = (v_1^{(\ell)}, \ldots, v_n^{(\ell)}))_{\ell = 1}^M$ is a sample of visible units obtained from the RBM with parameters $(W, \vec{b}, \vec{c})$. This estimator may suffer from high variance, especially for high values of $t^*$, which is unfortunately exactly the regime we are interested in. 

The AIS estimation procedure interprets the exponential term in Equation \eqref{eq:partitions_ratio} as the ratio of two unnormalized RBM distributions:

\begin{equation}
\label{eq:AIS_ratio}
    Z_{t^*}/Z \approx \frac{1}{M} \sum_{\ell = 1}^M \frac{\exp \left( t^* \sum_{i=1}^M v_i^{(\ell)} - E(\vec{v}^{(\ell)}, \vec{h}) \right)}{\exp \left( - E(\vec{v}^{(\ell)}, \vec{h}) \right)}
\end{equation}

where the energy at the numerator belongs to the tilted RBM with parameters $(W, \vec{b} + t^*\vec{u}, \vec{c})$, while the one at the denominator to the RBM with parameters $(W, \vec{b}, \vec{c})$. Indeed if evaluated on a Monte Carlo sampling from the RBM with parameters $(W, \vec{b}, \vec{c})$, the distributions of these two RBMs cannot in general be expected to be sufficiently close. This is the main reason of the high variance of the estimator.

\vspace*{0.5cm}
\begin{algorithm}[H]
\caption{Annealed Importance Sampling (AIS) estimation of $Z_{t^*}/Z$.}
\label{alg:AIS}
\KwData{$t^* \geq 0$; RBM with parameters $(W,\vec{b},\vec{c})$; $T$ temperatures, $(t_k)_{k=1}^T$, uniformly spaced on $[0, t^*]$, $M$ number of AIS runs.}
\KwResult{Estimate of $Z_{t^*}/Z$ with standard deviation.}
\For{$j = 1$ \KwTo M}{
    $\varepsilon^{(j)} \gets 1$\;
    \For{$k = 1$ \KwTo T}{
        \eIf{$k = 1$}{
        Sample $\vec{v}^k$ from RBM $(W, \vec{b}, \vec{c})$ with random initialization;
        }{
        Sample $\vec{v}^k$ from one-step Gibbs sampling of RBM $(W, \vec{b} + t_k\vec{u}, \vec{c})$ initialized with $\vec{v}^{k-1}$;
        }
        $\varepsilon^{(j)} \gets \varepsilon^{(j)} \cdot \exp\left( (t_k - t_{k-1}) \sum_{i=1}^n v^{k}_i \right)$
    }
}
Compute $\hat{\varepsilon} = \frac{1}{M} \sum_{j=1}^M \varepsilon^{(j)}$\;
Compute $\hat{\sigma}^2 = \frac{1}{M - 1} \sum_{j=1}^M (\varepsilon^{(j)} - \hat{\varepsilon})^2$\;
\Return{$\hat{\varepsilon}$ with standard deviation $\hat{\sigma}/\sqrt{M}$.}
\end{algorithm}
\vspace*{0.5cm}

The solution put forward by the AIS procedure is to define a sequence of intermediate RBMs with parameters $(W, \vec{b} + t_k \vec{u}, \vec{c})$, where $(t_k)_{k=1}^T$ form a grid on $[0, t^*]$ of suitable thickness. The parameters $(t_k)_{k=1}^T$ are also called \emph{temperatures} and allow us to interpolate smoothly from one distribution to the other.

The ratio in Equation \eqref{eq:AIS_ratio} can then be approximated by a telescoping product of ratios, each computed for a pair of consecutive RBMs in the temperature sequence, $(t_{k-1}, t_k)$, under the distribution of temperature $t_{k}$. In this way, the ratios are always computed on distributions that are very close to each other and the estimator is gradually tilted away from the original RBM to the fully tilted one, thus performing an annealed version of importance sampling. A complete description of the AIS estimation procedure can be found in Algorithm \ref{alg:AIS}. As the number of temperatures and the number of AIS runs increase, the accuracy of the estimator increases. In practice, good results can be obtained for $T \approx 20000$ and even low values of $M$, such as $M \approx 100$.

\subsection{Performance of the IS estimator}

We test the IS estimator in Equation \eqref{eq:IS_estimator} by training the credit RBM model on a synthetic dataset of default probabilities generated from a one-factor Gaussian copula portfolio comprising 250 obligors with heterogeneous average default probabilities (uniformly distributed between 2\% and 10\%) and with equicorrelation coefficient $\rho = 0.2$.

\begin{figure}[H]
    \centering
    \includegraphics[width=0.7\textwidth]{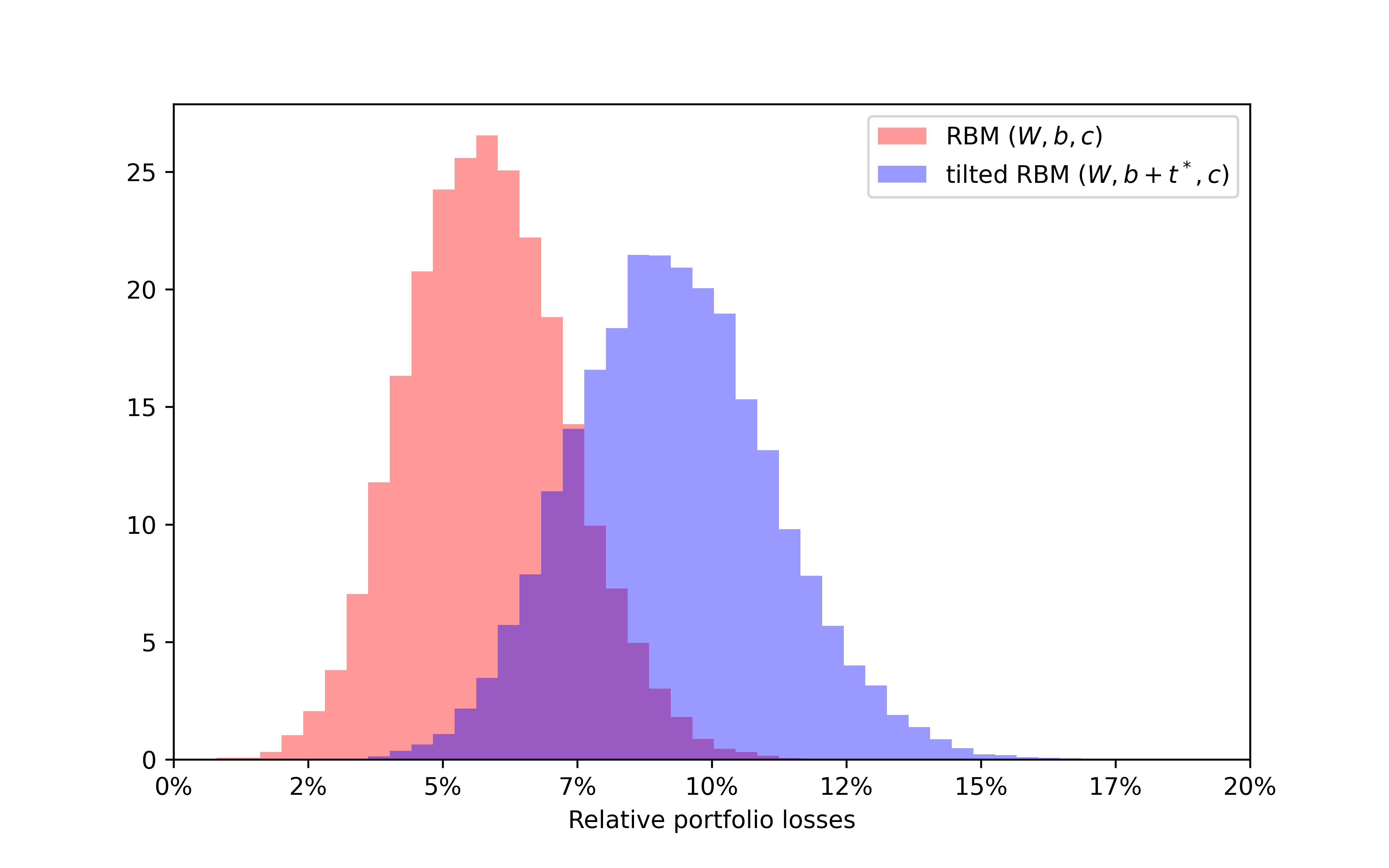}
    \caption{Histogram of relative portfolio losses for the original RBM and the tilted RBM.}
    \label{fig:IS_histogram}
\end{figure}

The optimal tilting, $t^*$, is chosen in such a way as to shift the mean relative portfolio loss from 6\% to approximately 9.5\%. This can be accomplished simply by trial and error, tilting the RBM of an arbitrary value $t$ and then evaluating the expected portfolio losses under this tilting via simple Monte Carlo sampling. For our choice of $t^*$ the histogram of the relative portfolio losses is shifted to the right as shown in Figure \ref{fig:IS_histogram}.

\begin{figure}[H]
    \centering
    \includegraphics[width=\textwidth]{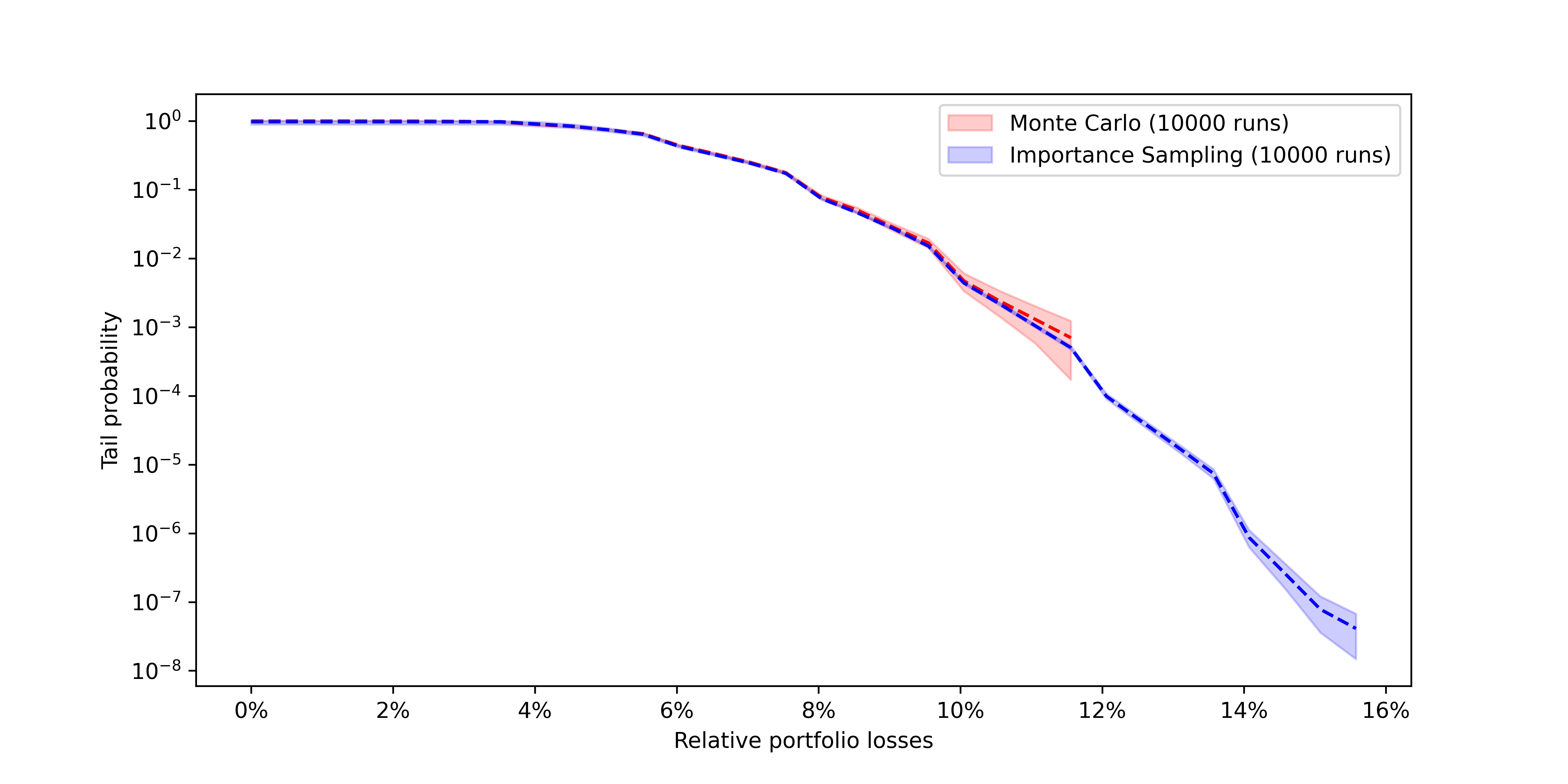}
    \caption{Estimated tail functions obtained by sampling from the RBM model using Monte Carlo simulations (in red) and using the importance sampling algorithm. Both estimators use 10000 samples. Colored shaded areas represent 95\% confidence intervals.}
    \label{fig:IS_tail}
\end{figure}

We can then compare the IS estimator in Equation \eqref{eq:IS_estimator} with a simple Monte Carlo sampling from an RBM that has not been tilted. The results are shown in Figure \ref{fig:IS_tail}. The two estimators agree on the initial portion of the tail, up to tail probabilities approximately of the order $10^{-3}$. Subsequently the variance of the simple Monte Carlo estimator starts diverging and the estimator itself vanishes, while the IS estimator, using the same number of samples, continues to map the tail function up to probabilities of order almost $10^{-8}$. Higher values of $t^*$ can be used to map the entire tail function to any degree of accuracy in a similar way, with no need to increase the number of simulations.

\section{Sector structure via hidden units' receptive fields}
\label{sec:sector_structure}

Latent factors in mixture models provide a low-dimensional representation of the covariates' dependence structure. The identification of latent factors that model dependencies between specific subgroups of covariates is of particular importance in finance, as it plays a role in portfolio diversification (by identifying sources of concentration risk) and in portfolio optimization (simplifying many operations, such as covariance matrix inversion \cite{de2018advances}). There is a rich literature in empirical finance concerned with the detection of sector structure from correlation matrices and its use in portfolio optimization \cite{de2016building, zhan2015application, bansal2004correlation, leon2017clustering}. We will show how to design similar procedures for the RBM model, by leveraging the specifics of its architecture.

In the credit RBM model the latent factors are extracted from data during training and are stored as random binary hidden units. In this section we see how latent factors can be extracted from these hidden units and be used to detect and monitor the sector structure of the credit portfolio.

We train a credit RBM model on a synthetic dataset\footnote{This section of the paper is based on synthetically generated data in order to validate our sector detection methodology. We need to know exactly the sector structure in order to make sure that the RBM recovers it correctly. Running the model on empirical data is a straightforward application.} of default probabilities generated from a high-dimensional multi-factor Gaussian copula model with 106 factors and 100 obligors, divided in five groups or \emph{sectors} of 20 obligors each. 

\begin{figure}[ht]
    \centering
    \includegraphics[width=0.75\textwidth]{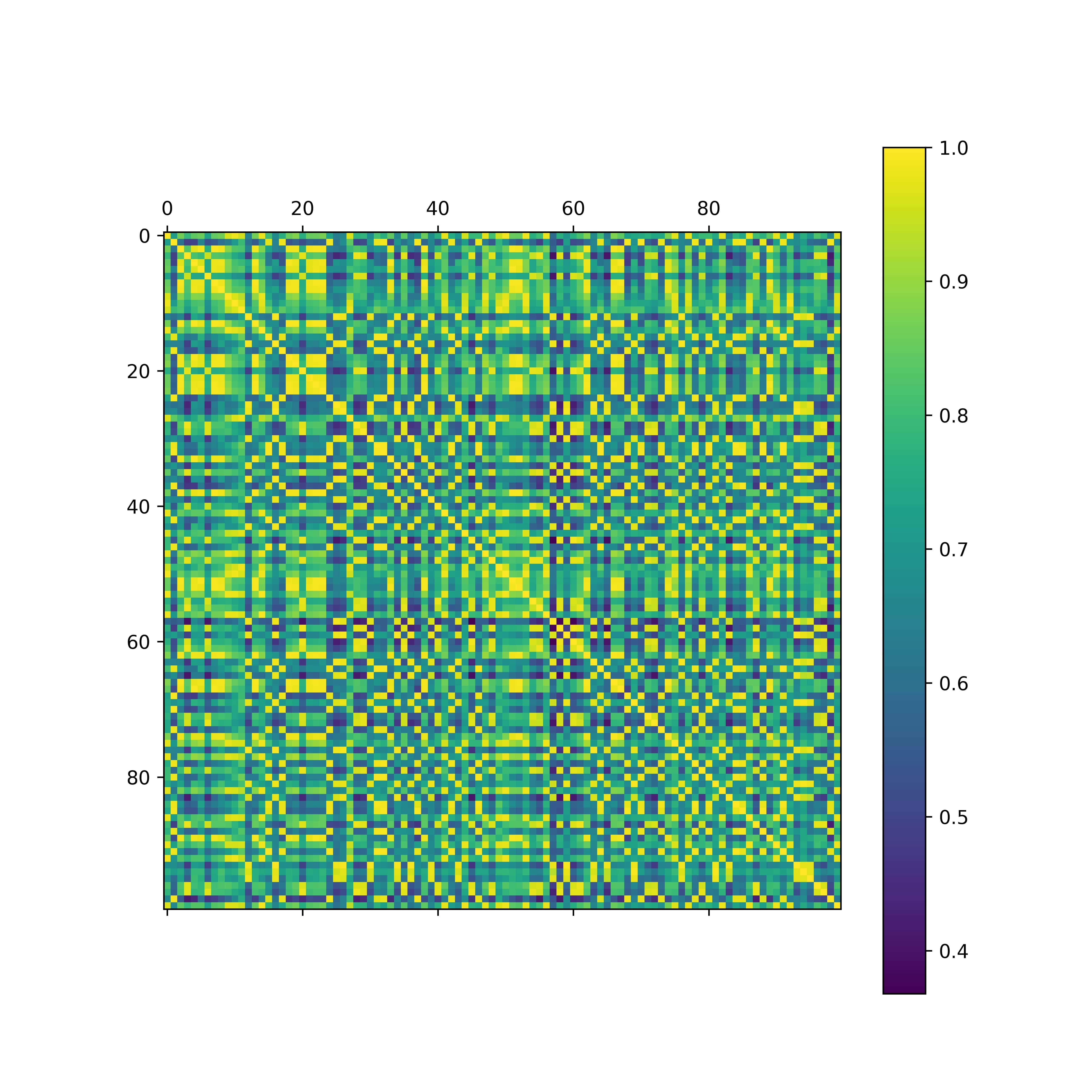}
    \caption{Correlation matrix of the artificially generated dataset with sector structure given by a multi-factor Gaussian copula.}
    \label{fig:corr_matrix}
\end{figure}

More specifically, the default probability of each obligor depends on three factors: a global factor, which is common to all obligors in the portfolio; a sector factor, which is shared among all obligors of the same sector; and finally an idiosyncratic factor, which is unique to each obligor. The sample correlation matrix of the default probabilities is shown in Figure \ref{fig:corr_matrix}, where all obligors have been shuffled, in order to hide the block structure of the matrix. Our goal is to perform correlation clustering using a trained RBM. In other words, we want to recover the way in which obligors are partitioned in sectors, so that when this partition is used to reorder the obligors in the sequence specified by its blocks, we can recover the correlation matrix in its original block configuration (up to permutations of the blocks).

The starting point is to analyze the so called \emph{receptive fields} of the hidden units, to borrow a term from biology. For each hidden unit we want to identify which visible units tend to activate it more often. In particular we are interested in activation patterns that result in higher defaults, i.e. higher values of the visible units themselves. 

From Equation \ref{eq:pds_in_RBM} we can readily see that the $j$-th hidden unit leads to higher default probabilities when $H_j = 1$ and $W_{ji} > 0$, as well as when $H_j = 0$ and $W_{ji} < 0$, with stronger impact for higher values of $|W_{ji}|$. For a given hidden unit, say the $j$-th hidden unit, we can therefore suppose that it encodes information about high default values of obligor $i$ if $|W_{ji}|$ takes a high value.

\begin{figure}[ht]
    \centering
    \includegraphics[width=\textwidth]{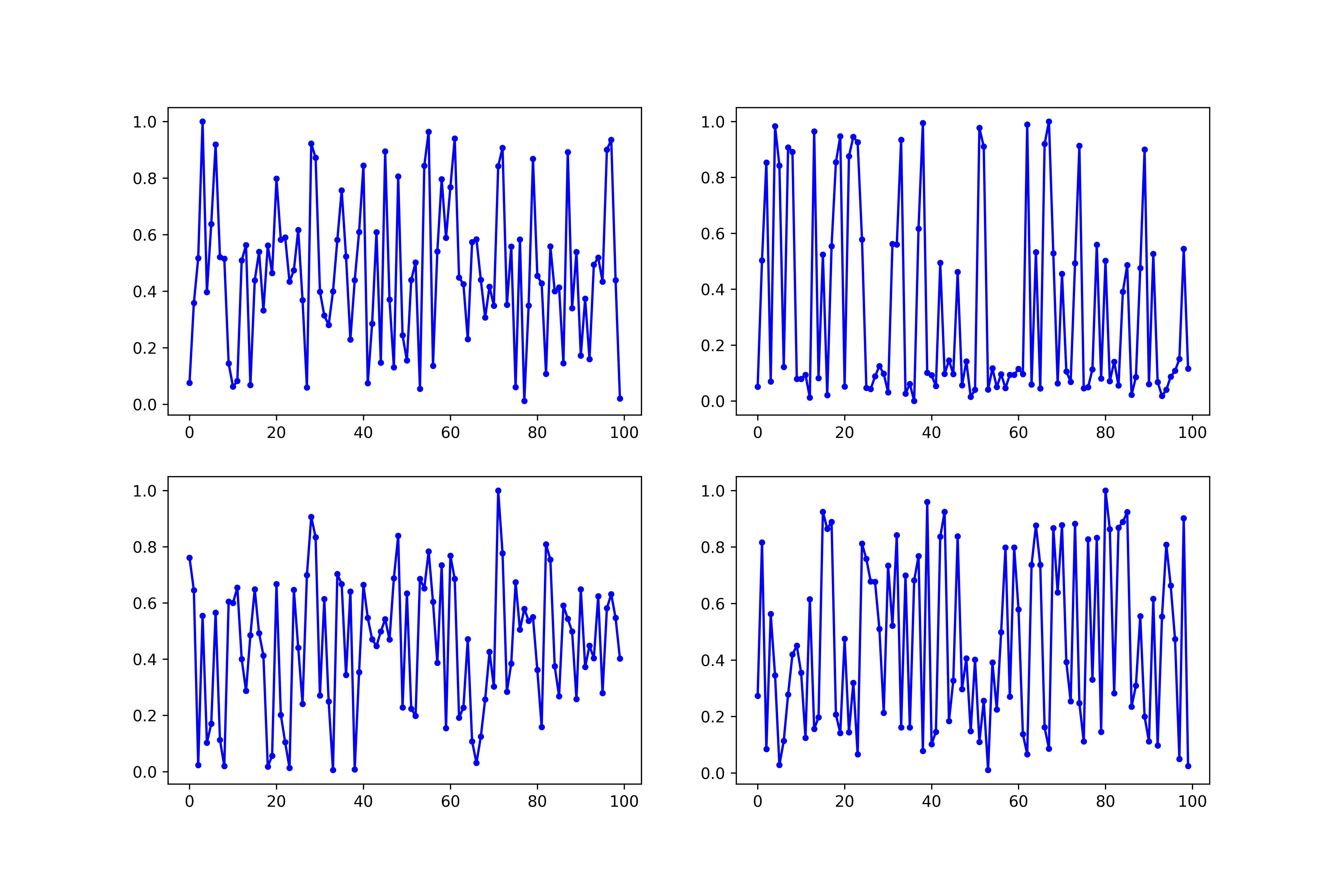}
    \caption{Plot of $\tilde{W}_{ji}$ for $i = 1, \ldots, n$ and four randomly selected hidden units.}
    \label{fig:receptive_fields}
\end{figure}

One way of displaying this quantity for the $j$-th hidden unit is to plot $|W_{ji}|$ as a function of $i = 1, \ldots, n$, normalized by its maximum value across all visible units, i.e. $\tilde{W}_{ji} := |W_{ji}|/\max_{i = 1, \ldots, n} |W_{ji}|$. Figure \ref{fig:receptive_fields} shows exactly these plots for four hidden units in the trained RBM. 

The hidden units clearly display a broad activation pattern, ranging from zero (i.e. no sensitivity to the corresponding visible unit) to one (maximum sensitivity), with a few selective activation peaks, that correspond to specific subgroups of visible units. In other words, each hidden unit appears to be specialized in modelling a local dependence, corresponding to the default of only few obligors.

These local dependencies can be graphically represented in an undirected, weighted graph, $G = (V, E)$, with vertex set $V = \{1, \ldots, n\}$, corresponding to the set of portfolio obligors. The edges of the graph are added according to the following rule: given a fixed threshold, $\varepsilon \in (0,1)$, we create an edge, $e=(l, k)$, between two vertices, $l$ and $k$, if there is at least one hidden unit, $j \in \{1, \ldots, m\}$, such that both $\tilde{W}_{jl}$ and $\tilde{W}_{jk}$ are above $1 - \varepsilon$. 

\begin{figure}[h!]
    \centering
    \includegraphics[width=0.6\textwidth]{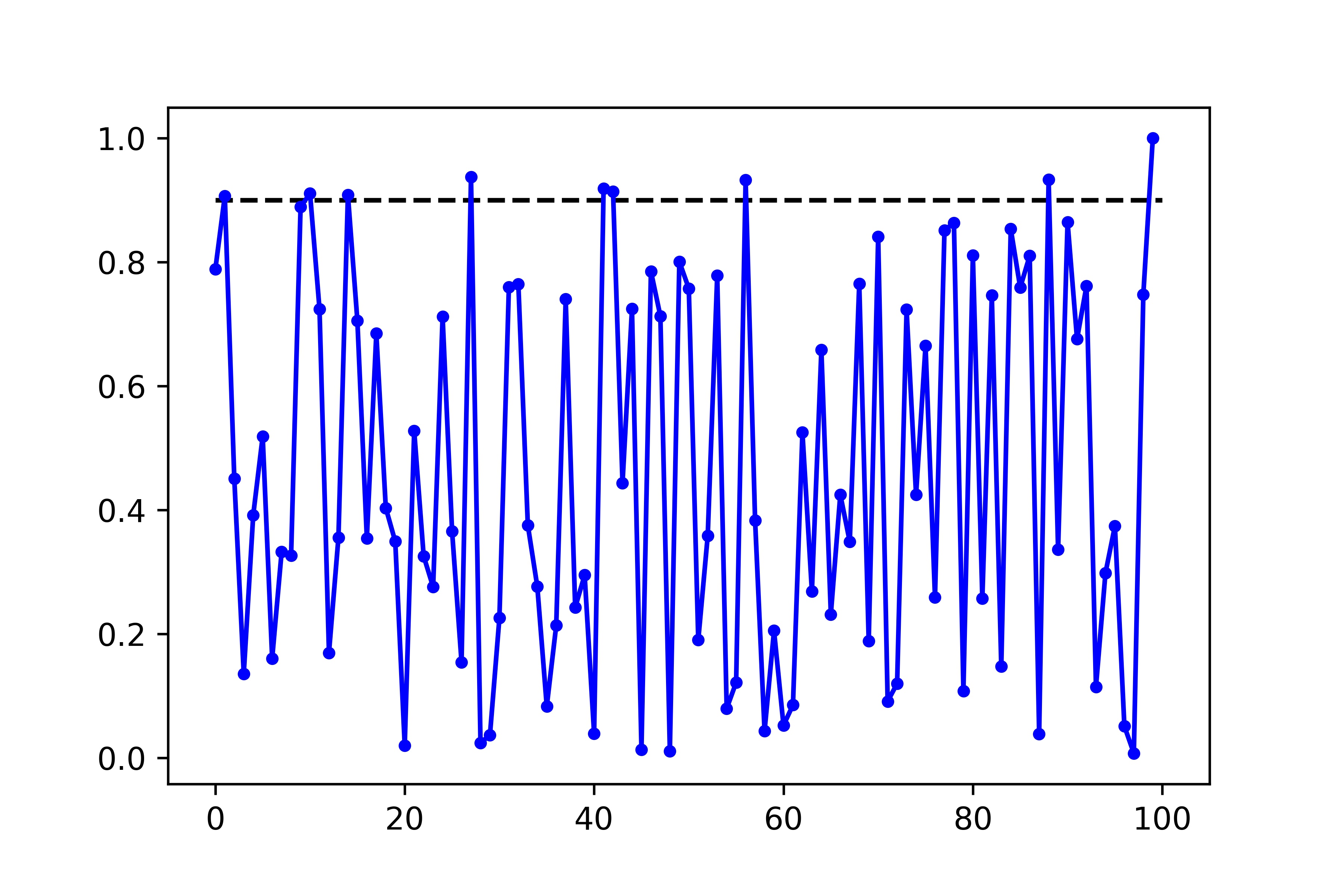}
    \caption{Thresholded plot of $\tilde{W}_{ji}$ for $i = 1, \ldots, n$ for $\varepsilon = 0.1$. The hidden unit is randomly chosen among the hidden units of the RBM.}
    \label{fig:receptive_field_thresholded}
\end{figure}

Graphically, for each hidden unit we threshold the values of $(\tilde{W}_{ji})_{i=1}^n$, as shown in Figure \ref{fig:receptive_field_thresholded}, and create edges between all pairs of visible units above this threshold. In this way each hidden unit induces a clique on the graph $G$.

Of course two different hidden units can give rise to the same edge. This information gets stored in the edge weight. More specifically, the number of hidden units that give rise to the same edge are counted in the edge weight, i.e. if $e=(l,k) \in E$, then 
$$w_e = |\{ j \in \{1, \ldots, m\} \:\: : \:\: \tilde{W}_{jl}, \tilde{W}_{jk} > 1 - \varepsilon \}|.$$ 
This implies that visible units that belong to the receptive fields of many hidden units are connected by edges with higher weight.

\begin{figure}
\centering
\begin{subfigure}{\textwidth}
\centering
  \includegraphics[width=0.45\linewidth]{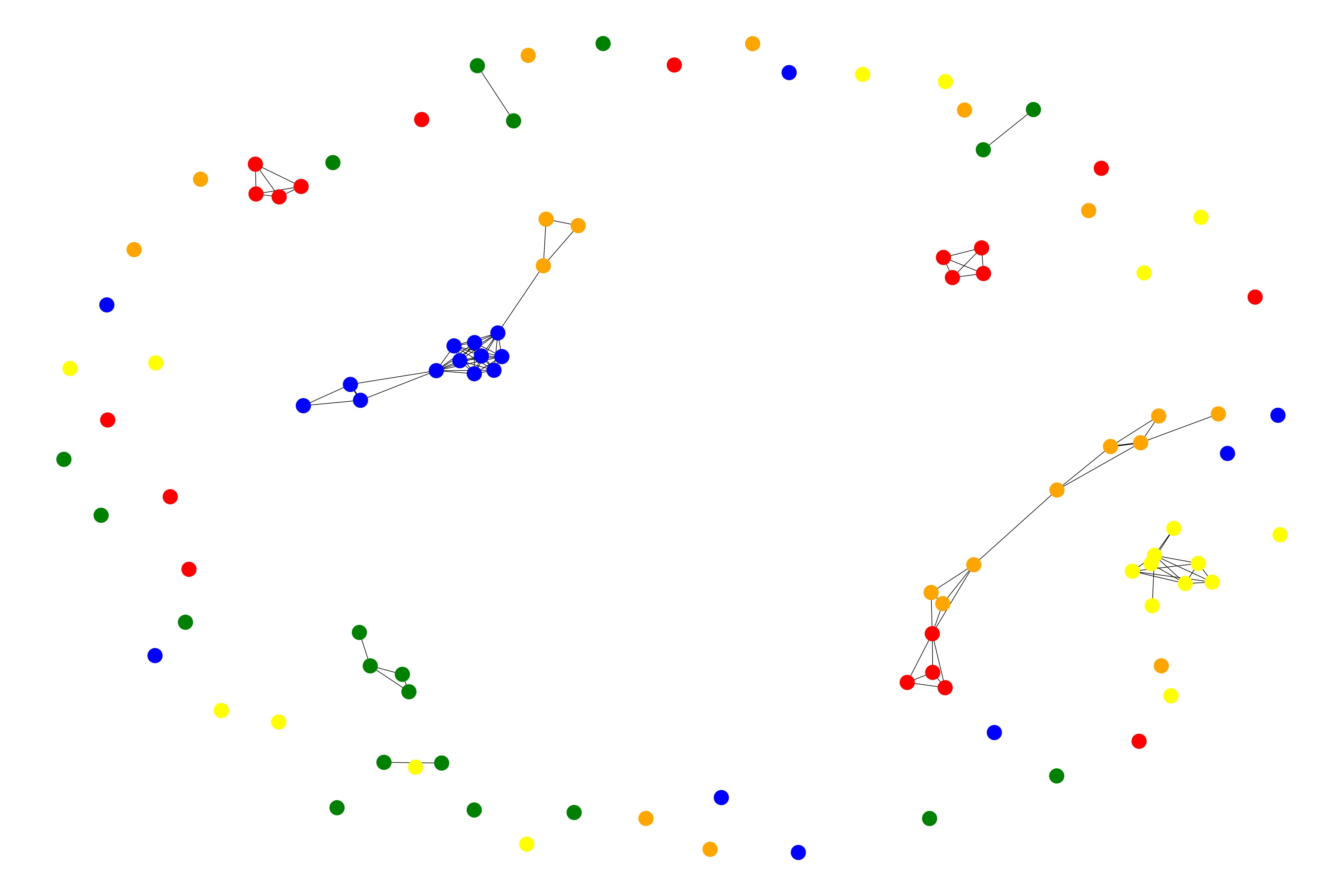}
  \subcaption{$\varepsilon = 0.03$}
\end{subfigure}
\begin{subfigure}{0.49\textwidth}
  \includegraphics[width=\linewidth]{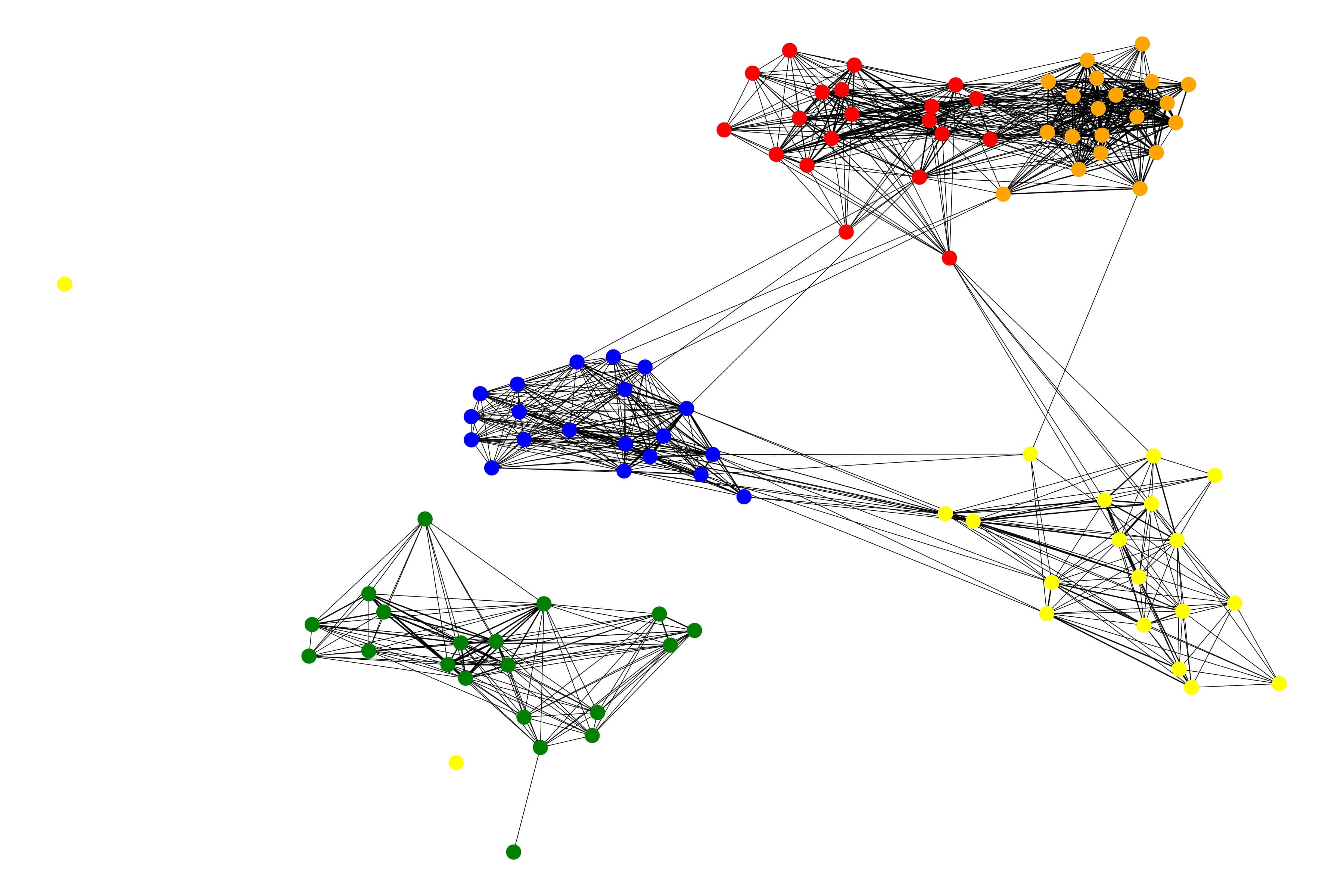}
  \subcaption{$\varepsilon \approx 0.075$}
\end{subfigure} %
\begin{subfigure}{0.49\textwidth}
  \includegraphics[width=\linewidth]{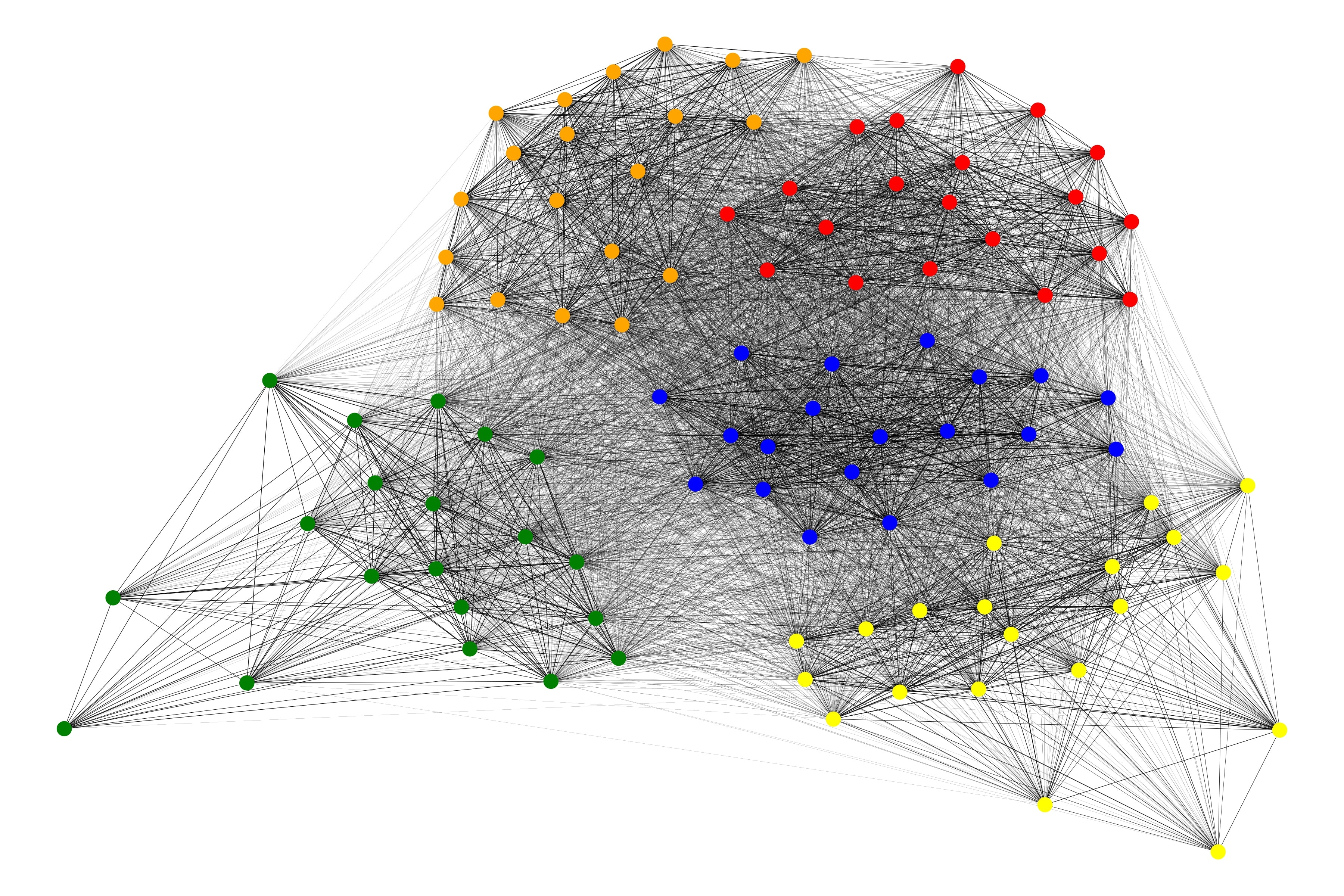}
  \subcaption{$\varepsilon = 0.25$}
\end{subfigure} %
\caption{Representations of $G$ for three values of the thresholding parameter $\varepsilon$. Nodes are colored according to their sector under the true data model. The graph layout is computed using the Fruchterman-Reingold force-directed algorithm \cite{fruchterman1991graph}.}
\label{fig:sector_structure_graphs}
\end{figure}

As the threshold value $\varepsilon$ is varied, the graph $G$ will in general display different topologies. Figure \ref{fig:sector_structure_graphs} shows three possible choices of $\varepsilon$ and the corresponding graphs. The nodes have been colored according to their sector under the true data model, which is an information that is not explicitly available to the RBM, nor to our graph generating algorithm. The graphs have a spring layout, which means that nodes connected by edges with higher weights are pulled together and appear closer in the graph. 

For low values of $\varepsilon$, as in figure (A), we fail to capture the full receptive field of most hidden units, so that the resulting graph is highly disconnected and the many connected components do not form sectors. 

For intermediate values (see (B)) we see that the graph is able to identify the true sector structure of the data, since all nodes of the same color form tightly connected components, with only few and peripheral connection across sectors. 

For high values of $\varepsilon$ (see (C)), instead, the local nature of the hidden units' receptive field is lost as we include more and more global correlations and the sectors become more intertwined.

\begin{figure}[h!]
    \centering
    \includegraphics[width=0.6\textwidth]{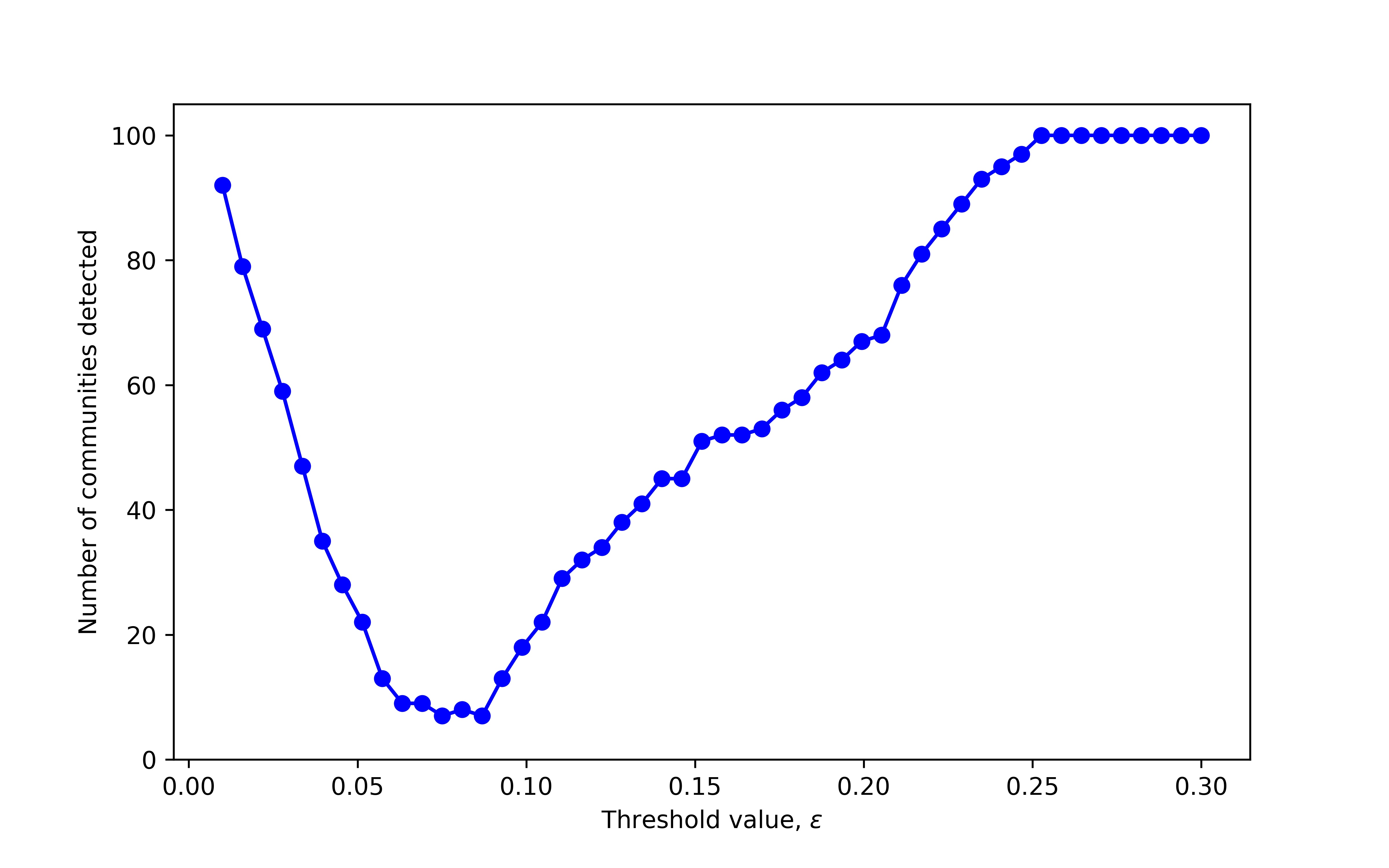}
    \caption{Number of communities detected on $G$ using greedy modularity maximization, as a function of the threshold parameter, $\varepsilon$.}
    \label{fig:number_of_communities}
\end{figure}

The graphical representations shown in Figure \ref{fig:sector_structure_graphs} are certainly useful and would already allow portfolio managers to identify the main local dependencies (i.e. sectors) in their portfolio, but in order to perform correlation clustering, we must be able to eventually extract a unique optimal partition of the vertices in sectors.

The identification of such a sector partition in $G$ can be fully automated by using any community detection algorithm for weighted graphs. One natural choice is the well-known Newman's greedy modularity maximization algorithm \cite{newman2004fast, newman2004analysis}, of which a version, the Clauset-Newman-Moore's algorithm \cite{clauset2004finding}, has the added benefit of being very efficient also on extremely large graphs ($\approx 10^7$ vertices). These algorithms detect communities of tightly connected vertices and return as final output a partition of the vertices such that, loosely speaking, edges within the partition blocks are maximized and edges across partition blocks are minimized. 

In Figure \ref{fig:number_of_communities} we show the number of sectors detected via greedy modularity maximization on the graph $G$ as $\varepsilon$ is varied. By its very definition, modularity tends to result in many communities both for highly disconnected graphs and for highly connected ones, so that the number of sectors as a function of $\varepsilon$ has a global minimum and allows the identification of an optimal threshold, corresponding to the minimum number of sectors. 

On our dataset this minimum number of sectors is seven and is attained at $\varepsilon \approx 0.075$. The corresponding graph was already shown in Figure \ref{fig:sector_structure_graphs} (B) and it correctly identifies all five sectors, with the exception of the addition of two spurious sectors corresponding to two isolated yellow vertices (left and lower left of the image).

\section{Stress testing via conditional Gibbs sampling}
\label{sec:stress_testing}

Stress testing is a risk management tool, adopted by financial institutions and regulatory authorities, 
which is used to assess the impact of a set of adverse macroeconomic scenarios on the performance of an institution's asset portfolio \cite{pop2017stress}. 

The first use of stress testing in banking regulation dates back to the Basel I accord \cite{basel1996basel}, in which large banking institutions were required to conduct regular stress tests by simulating losses on their trading book under adverse macroeconomic scenarios, of a nature and magnitude similar to historically important market shocks. With Basel II \cite{basel2004basel} stress testing techniques started being applied also to risks other than market risk and played an important role for the determination of capital requirements. Finally, after 2007 several regulatory bodies began mandating regular stress test exercises. In 2012 the Federal Reserve Board (FRB) consolidated old practices into a new framework, which nowadays comprises the annual Dodd-Frank Act stress test and the Comprehensive Capital Analysis and Review (CCAR), the results of which are regularly released to the public and constitute a cornerstone of the supervisory activity of the FRB. 

\begin{figure}[H]
    \centering
    \includegraphics[width=0.75\textwidth]{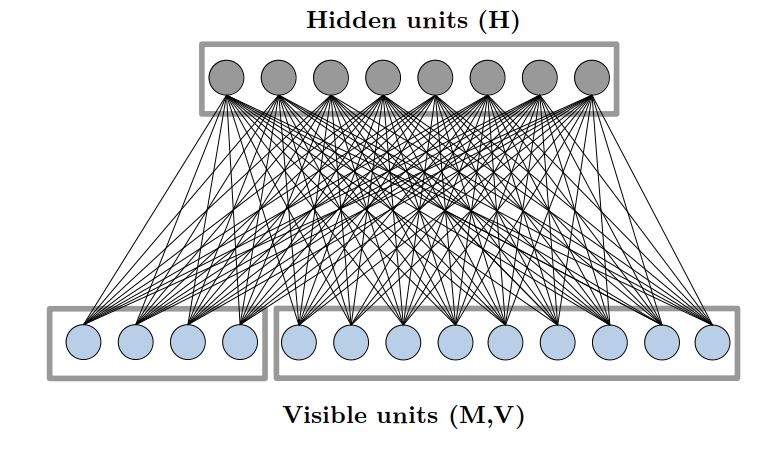}
    \caption{Restricted Boltzmann Machine for stress testing. The visible units consist of macroeconomic variables, $\vec{M}$, and portofolio losses, $\vec{V}$.}
    \label{fig:RBM_stress_test}
\end{figure}

In this section we will focus on the FRB's implementation of the Dodd-Frank Act stress testing methodology, which is paradigmatic of virtually all stress testing methodologies implemented worldwide. Under this methodology a set of three macroeconomic scenarios is produced by independent economists: the baseline scenario, the severely adverse scenario, and the alternative severe scenario. Each scenario is specified in terms of 28 macroeconomic variables, divided in 16 domestic variables (e.g. US real and nominal GDP, US unemployment rate and disposable income, etc.) and 12 international variables (e.g. foreign currency exchange rates, real and nominal GDP growth in four different world areas, etc.). The individual financial institutions taking part in the exercise are then required to use their asset-specific, internal risk models (e.g. credit risk, market risk, and liquidity risk models) to assess the impact of each macroeconomic scenario on their overall portfolio profits and losses. 

Mathematically speaking, if we denote by $\vec{M} = (M_1, \ldots, M_k)$ the vector of macroeconomic variables used for scenario specification, a stress test exercise on a credit portfolio amounts to the evaluation of the conditional distribution of the total portfolio losses, $L_n$, given $\vec{M}$. More specifically, one is interested in evaluating the conditional tail function, $\P{L_n > x|\vec{M}}$, or the various conditional risk measures, such as, for instance, the conditional Value at Risk, $\text{VaR}_\alpha(L_n|\vec{M})$, or even the conditional default probabilities of individual borrowers in a portfolio, $\P{V_i = 1|\vec{M}}$.

The joint probability distribution of macroeconomic factors and portfolio losses can be learned by training an RBM with visible units $(\vec{M}, \vec{V})$ on a joint dataset of historical time series (see Figure \ref{fig:RBM_stress_test}). Once the training of the RBM is successfully completed, any functional of the distribution of $\vec{V}$ given a particular macroeconomic scenario $\vec{M} = (m_1, \ldots, m_k)$ can be estimated by conditional blocked Gibbs sampling, which corresponds exactly to a standard blocked Gibbs sampling procedure, as in Algorithm \ref{alg:Gibbs_sampling}, with the exception that the visible units corresponding to the macroeconomic variables, $\vec{M}$, are clamped to the values corresponding to the relevant scenario, $(m_1, \ldots, m_k)$, at each iteration. As this Gibbs sampling is performed for sufficiently many steps, the corresponding Markov chain will converge to the conditional distribution of $\vec{V}$ given the scenario $\vec{M} = (m_1, \ldots, m_k)$.


To illustrate the kind of analyses made possible by this procedure, we implement the FRB's Dodd-Frank stress test of June 2020 on the same empirical credit portfolios we used in Section \ref{sec:estimation_of_risk_measures} (see Appendix \ref{sec:empirical_data} for more information), augmented by the historical time series of quarterly values of the reference macroeconomic variables used by the FRB\footnote{These historical quarterly time series are available on the FRB website \texttt{https://www.federalreserve.gov/supervisionreg/dfa-stress-tests.html}. We obtained daily data from quarterly data by interpolating in a non-anticipating way (i.e. constant interpolation from last known value). This interpolation method was preferred to linear (or more complex) interpolation techniques, because it does not inject information about future quarterly values at previous time steps.}. 

\begin{figure}[H]
    \centering
    \includegraphics[width=\textwidth]{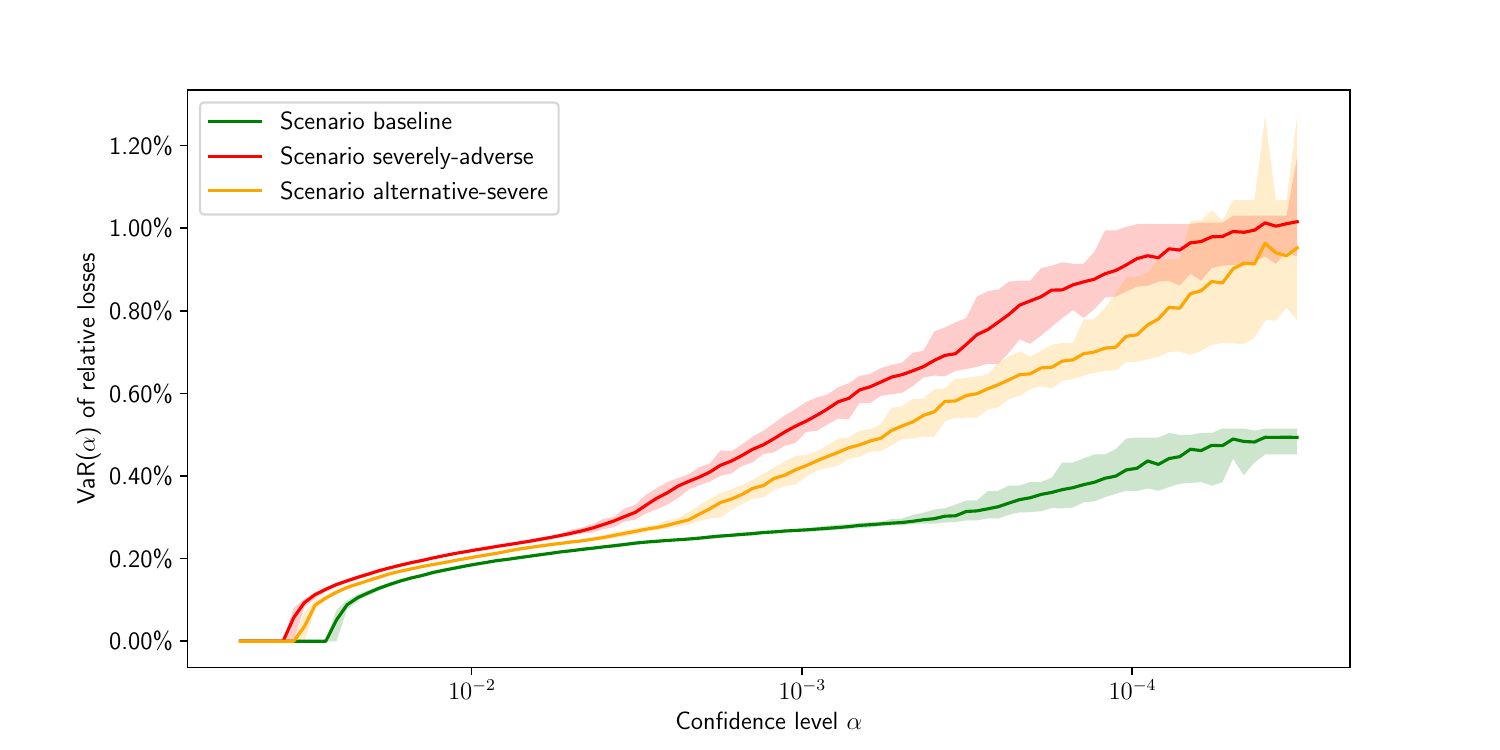}
    \caption{Estimation of stressed Value at Risk of portfolio with A rating for different confidence levels and FRB's 2020Q4 stress test scenarios. The colored areas represent 95\% non-parametric bootstrap confidence bands (100'000 Monte Carlo samples, 100 bootstrap samples).}
    \label{fig:stress_test_vars}
\end{figure}

Conditional sampling from the RBM allows also the computation of stressed risk measures, such as the Value at Risk depicted in Figure \ref{fig:stress_test_vars} for our A-rated portfolio, under different scenarios. Both the severely-adverse and alternative-severe scenarios determine a clear shift to the right towards higher losses, with respect to the baseline scenario. The two riskier scenarios determine a statistically significant increase of the VaR at almost all confidence levels and the credit RBM model can be used to provide a quantitative estimate (with associated statistical uncertainty) of the additional capital required to cover the losses under such macroeconomic scenarios.

The results presented in this section are qualitatively similar across all rating classes in our dataset. See Section \ref{subsec:stressed_var_all_rating_classes} for an overview.

\section{Conclusions}

In this paper we have introduced a credit risk model based on Restricted Boltzmann Machines (RBMs) and investigated its performance across several financial tasks. RBMs are universal approximators for loss distributions and can thus provide a non-parametric alternative to the commonly used copula models, which are known to generate substantial model risk due to their over-reliance on parametric assumptions. 

At the same time the model is shown to retain mathematically appealing properties, such as a conditional independence structure and an explicit formula for the model distribution, which make it amenable to many of the same mathematical techniques known for classical credit risk mixture models.

We have shown on an empirical dataset that the credit RBM model provides better fits of the empirical loss distribution than a multi-factor Gaussian or t copula model and that, in particular, it provides better estimates for risk measures.

We have also introduced a novel importance sampling estimator for the tail probabilities of total portfolio losses under the model, thus showing that risk measures can be computed efficiently and accurately to any confidence level. This is a substantial advantage with respect to most other credit risk models, which typically must rely on computationally expensive Monte Carlo simulations.

It has been shown that the model can also be used for automated correlation clustering and sector detection, so that the latent factors extracted by the RBM model can be interpreted in terms of local dependencies among obligors. This could provide practitioners with an important tool for portfolio management and identification of concentration risk.

Finally, we have seen how the model naturally lends itself to the implementation of portfolio stress tests, by simulating three macroeconomic scenarios from the FRB's June 2020 Dodd-Frank Act stress test on an empirical portfolio.

\appendix

\section{Dataset description}
\label{sec:empirical_data}

The full dataset consists of daily one-year default probabilities for 1'012 publicly listed US companies from 4 January 1999 to 30 December 2022. 

The default probabilities come from the Bloomberg Corporate Default Risk Model (DRSK) for Public Firms (see \cite{bondioli2021bloomberg} for a description), which is a proprietary structural model loosely based on Merton's firm-value model \cite{merton1974pricing}. The goal of Bloomberg's DRSK is to provide accurate estimates of the physical default probability of corporate entities, which can be used for risk management purposes (e.g. determination of capital requirements). We emphasize that our credit RBM model can also be used to estimate risk-neutral default probabilities, by training, for instance, on market-implied default probabilities from single-name CDS spreads (or any other risk-neutral estimate of default probabilities). In this sense, the particular choice of dataset done in this paper is arbitrary and it only serves the purpose of showcasing the model.

\begin{figure}[H]
    \centering
    \includegraphics[width=\textwidth]{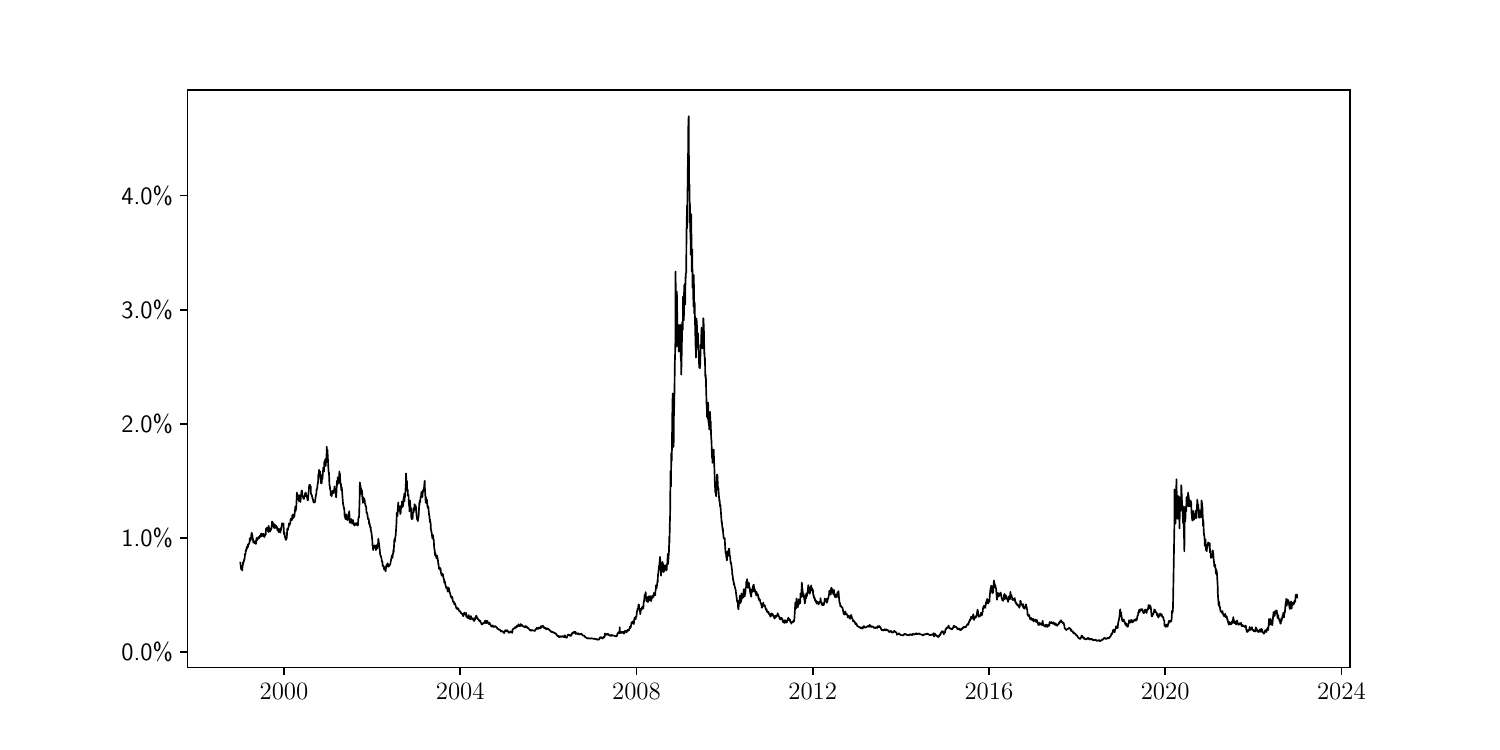}
    \caption{Time series of average default probability in the full dataset.}
    \label{fig:mean_pds}
\end{figure}

Figure \ref{fig:mean_pds} shows the time series of the average default probability in the dataset. The peaks in default probability correspond to big economic and/or financial shocks, such as the September 11 attacks, the Venezuelan oil strike (end 2002, beginning 2003), the financial crisis of 2008 and the onset of the Covid pandemic in 2020. The average default probability in the dataset across time and all firms is approximately 0,6\%.

The companies have been selected from the Russell 3000 index, which is an index covering 97\% of the US equity market capitalization. In order to maximize the number of datapoints in time, only firms present in the Bloomberg DRSK dataset from its inception have been included, resulting in our final selection of 1'012 firms. 

\begin{figure}[H]
    \centering
    \includegraphics[width=\textwidth]{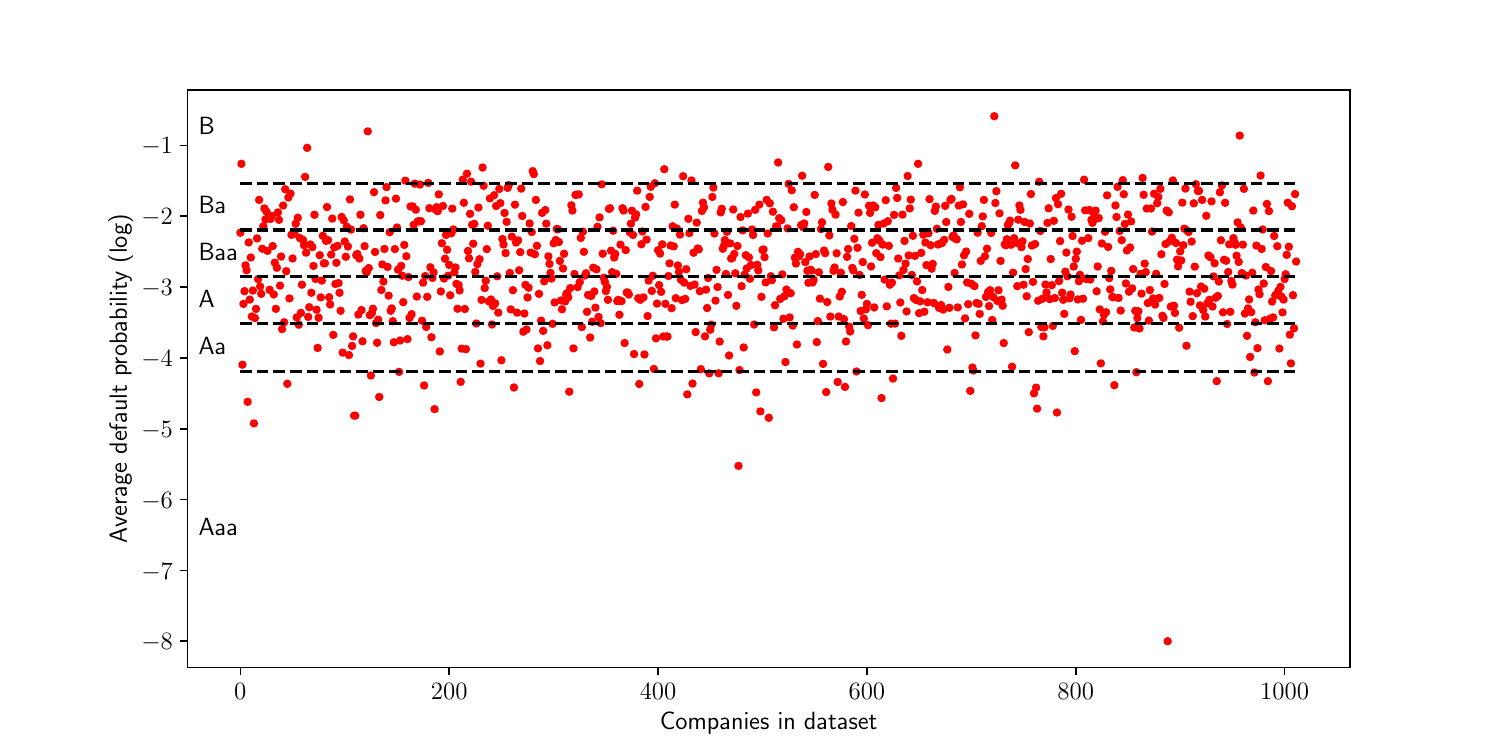}
    \caption{Breakdown of dataset by average default probability and rating class.}
    \label{fig:rating_classes}
\end{figure}

This selection covers the full range of corporate credit ratings, as shown in Figure \ref{fig:rating_classes}, where we plotted the average default probability in time for each company in the dataset, superimposed on a grid of Moody's corporate bond rating buckets\footnote{Moody's Investors Service does not provide an official mapping between their ratings and default probability buckets, but it is common practice to estimate such a mapping from historical default data. Here we used the mapping suggested in \cite[Figure 1.1]{bluhm2016introduction}. We further aggregated credit ratings by letter (e.g. ratings Aa1, Aa2 and Aa3 have been aggregated in the rating class Aa).}.

In order to investigate how our results generalize across credit ratings, we have subdivided the firms in our dataset into six homogeneous credit portfolios according to the buckets shows in Figure \ref{fig:rating_classes}. Table \ref{tab:rating_classes} provides a summary overview of each portfolio, while Tables \ref{tab:tickers1} and \ref{tab:tickers2} provide the complete list of tickers.

All the results of this paper have been tested independently for these six portfolios and have been reported either in the main text or in the appendices. Whenever results varied appreciably across rating classes, we have mentioned it explicitly in the main text.

\begin{table}[H]
    \centering
    \begin{tabular}{l|c|c}
         \textbf{Credit rating} & \textbf{Number of companies} & \textbf{Default probability bucket} \\ \hline
         Aaa & 39 & [0.000\%, 0.007\%]\\
         Aa & 88 & [0.007\%, 0.031\%]\\
         A & 302 & [0.031\%, 0.139\%]\\
         Baa & 316 & [0.139\%, 0.638\%]\\
         Ba & 238 & [0.638\%, 2.927\%]\\
         B & 29 & [2.927\%, 100\%] \\ \hline
         \textbf{Total} & 1012
    \end{tabular}
    \caption{Breakdown of dataset by average default probability and rating class.}
    \label{tab:rating_classes}
\end{table}

\begin{table}[H]
    \centering
    \begin{tabular}{l|p{12cm}}
         \textbf{Credit rating} & \textbf{Tickers} \\ \hline
         Aaa & \tiny ABT, ADP, AMGN, BF/A, BF/B, BMY, BRO, CASS, CHD, CL, COST, DHR, EXPD, GIS, HRL, HSY, IDXX, INTC, JNJ, KMB, KO, LANC, MDT, MKC, MMM, MRK, NATI, NEOG, PEP, RLI, RMD, RNR, SAM, SSD, SYK, TR, UTMD, WDFC, WMT\\
         Aa & \tiny AAON, AMAT, AOS, ATR, AZO, BAX, BCPC, BDX, BEN, BIO, BRK/B, CAH, CASY, CBSH, CHE, CINF, CLX, CPB, CPRT, CTAS, CVX, DCI, DOV, DRQ, EBAY, ECL, EFX, EL, EMR, FAST, FCNCA, FFIN, GD, GGG, GPC, GWW, HAE, HD, HEI, HEI/A, HUBB, HWKN, IBM, IEX, IFF, INTU, ITW, JOE, K, KLAC, LECO, LLY, LOW, LSTR, MCD, MCY, MMS, MNST, MO, MTD, NEE, NFG, NKE, ORCL, PFE, PG, PGR, PSA, QCOM, RHI, ROG, ROL, ROP, RTX, SEIC, SNPS, SXT, TAP, TTC, TXN, VICR, WABC, WAT, WERN, WST, XOM, XRAY, ZBRA \\
         A & \tiny AAPL, ABC, ACNB, ADI, ADM, ADSK, AEE, AEP, AFG, AFL, AIT, ALB, ALE, ALL, AME, AMNB, AON, APD, APH, AROW, ARTNA, ASB, ATNI, ATO, ATRI, ATVI, AUB, AVY, AWR, AZZ, BAC, BALL, BANF, BHB, BIIB, BMI, BOH, BOKF, BRC, BRKL, BSX, BWA, CAC, CACI, CADE, CAG, CB, CBT, CBU, CCF, CDNS, CFR, CHCO, CHH, CIVB, CMA, CMCSA, CNOB, COLB, COLM, COO, COP, CPK, CSL, CTBI, CTO, CTRA, CVBF, CVS, CW, CWT, D, DGX, DHIL, DIS, DOX, DRI, DTE, DUK, EBF, ED, EGP, EMN, EOG, ES, ESE, ETN, ETR, EXC, EXP, EXPO, FCBC, FCF, FELE, FFBC, FFIC, FHB, FHI, FICO, FISV, FIZZ, FLIC, FMC, FMNB, FNB, FNLC, FORR, FULT, FWRD, GABC, GBCI, GEF, GEF/B, GEN, GILD, GL, GNTY, GRC, GSBC, GTY, HAIN, HAS, HBNC, HE, HIFS, HLIO, HOLX, HON, HP, HPQ, HSIC, HTLF, HWC, IBOC, IDA, IDCC, INDB, INT, ITT, JJSF, JPM, KEX, KR, L, LCII, LEG, LFUS, LHX, LKFN, LMT, LNN, LNT, MAR, MATW, MATX, MCHP, MCO, MDU, MGEE, MKL, MLI, MLM, MMC, MSA, MSEX, MSM, MTB, MTCH, MTX, NBN, NBTB, NDSN, NEM, NJR, NOC, NTAP, NUE, NUS, NVR, NWBI, NWN, NYCB, OCFC, OGE, OLED, OMC, ONB, ORI, ORLY, OZK, PDCO, PEBO, PEG, PFC, PGC, PH, PII, PKE, PNC, PNR, PNW, POOL, POWI, POWL, PPG, PPL, PRGO, PRK, RBCAA, RES, RGA, RL, RNST, ROK, ROST, RPM, RRX, RSG, RYN, SASR, SBSI, SBUX, SCCO, SCL, SCSC, SFNC, SGC, SHW, SIGI, SIVB, SJW, SLB, SMBC, SNA, SO, SON, SPGI, SR, SRCE, SRE, SSB, STBA, STE, STZ, SWK, SWX, SYBT, SYY, T, TCBK, TFC, TFX, TG, TGT, THFF, TJX, TMO, TMP, TNC, TRMK, TRST, TRV, UBA, UBSI, UEIC, UGI, UHS, UHT, UMBF, UMH, UNF, UNH, UNP, USB, UTL, UVSP, UVV, VFC, VLY, VMC, VMI, VZ, WAFD, WASH, WBA, WEC, WEYS, WFC, WIRE, WLY, WOLF, WPC, WRB, WSBC, WSFS, WSO, WTFC, WTM, WTRG, WTS, WWD, YUM \\
    \end{tabular}
    \caption{Constituents of credit portfolios in dataset (ratings: Aaa, Aa, A).}
    \label{tab:tickers1}
\end{table}

\begin{table}[H]
    \centering
    \begin{tabular}{l|p{12cm}}
         \textbf{Credit rating} & \textbf{Tickers} \\ \hline
         Baa & \tiny AA, ABCB, ABM, ACIW, ADC, AEIS, AGCO, AIN, AKR, ALCO, ALG, ALX, AMG, AMSWA, AMWD, AMZN, AN, APA, APOG, ARCB, ARE, ARW, ASH, ASTE, ATRO, AVA, AVB, AVD, AVID, AVT, AXP, AZTA, B, BA, BBWI, BBY, BCO, BFS, BH, BHE, BK, BKE, BKH, BLX, BPOP, BUSE, BXP, C, CACC, CAKE, CASH, CAT, CATY, CBAN, CBRL, CHDN, CI, CIX, CMC, CMI, CMS, CMTL, CNA, CNMD, CNP, COHR, COKE, CPT, CRAI, CRS, CSGS, CSR, CSX, CTS, CZNC, DCO, DDD, DE, DGII, DHI, DIOD, DISH, DLTR, DLX, DVN, DX, DY, EAT, ELME, ELS, EME, ENTG, EPAC, EQR, EQT, ESS, ETD, FBNC, FCN, FCX, FDP, FDX, FE, FHN, FITB, FLS, FRBK, FRME, FRPH, FRT, FSS, FUL, GATX, GE, GEO, GFF, GLT, GLW, GPS, GVA, HAL, HBAN, HCKT, HELE, HES, HIBB, HIW, HMN, HNI, HOG, HRB, HSC, HTBK, HUM, HVT, IMKTA, INGR, IP, IPAR, IPG, IRM, ITRI, JACK, JBHT, JBL, JCI, JEF, JLL, JWN, KAI, KAMN, KBAL, KELYA, KEY, KIM, KMPR, KMT, KRC, KSS, KWR, LAMR, LBAI, LH, LRCX, LSI, LTC, LUMN, LUV, MAA, MAN, MAS, MAT, MATV, MBWM, MCK, MD, MDC, MFA, MGRC, MHK, MIDD, MLKN, MMSI, MNRO, MODG, MOG/A, MOV, MRO, MS, MSI, MTN, MTRN, MU, MUR, MYE, MYGN, NATR, NCR, NHI, NI, NL, NLY, NNN, NOV, NSC, NSIT, NTRS, NWL, NWPX, NYT, O, ODFL, OFC, OFG, OFIX, OKE, OLN, OMI, ONTO, OSK, OSPN, OTTR, OXM, OXY, PCAR, PCH, PEAK, PHM, PKI, PLXS, PNM, POR, PRG, PSMT, PTC, PTEN, PXD, QCRH, R, RAMP, REG, REGN, RF, RGEN, RJF, RMBS, ROCK, RS, SBCF, SCHL, SCHN, SCHW, SEB, SEE, SENEA, SF, SHOO, SHYF, SKT, SMG, SNV, SPB, SPXC, SRCL, STC, STLD, SUI, SWKS, SXI, TBI, TDS, TER, TGNA, THG, TKR, TOL, TRMB, TRN, TSCO, TSN, TT, TTEC, TTEK, TTWO, TXT, TYL, UDR, UFPI, UNM, UNTY, USM, USPH, VECO, VGR, VIAV, VLGEA, VLO, VNO, VRE, VRTX, VSAT, VSH, VTRS, VVI, WAB, WBS, WEN, WHR, WM, WOR, WRLD, WSM, WWW, WY, XEL, ZION \\
         Ba & \tiny AES, AGR, AGX, AGYS, AIG, AIR, AIV, AJRD, ALK, ALKS, AMD, AMED, AMKR, AMOT, ANDE, AORT, ATI, AVNT, AXTI, BANR, BBSI, BC, BCRX, BDN, BIG, BJRI, BOOM, BRT, BYD, BZH, CAL, CALM, CBZ, CCK, CCL, CDE, CDMO, CENT, CENTA, CENX, CERS, CHS, CLDX, CLF, CLFD, CLH, CMCO, CNTY, COF, CPF, CRD/A, CRMT, CRUS, CSV, CTLP, CUZ, CVLG, CWST, CXW, DAR, DBD, DDS, DECK, DEN, DENN, DIN, DO, DORM, DVA, EAF, EEFT, EHC, EIX, EPR, EQC, EZPW, F, FBP, FC, FCEL, FIX, FL, FLL, FOSL, FR, GBX, GCO, GERN, GES, GHC, GIC, GIII, GPI, GT, GTN, HA, HIG, HL, HLIT, HRTX, HUBG, HXL, HZO, IBCP, IIIN, IMAX, INVE, IONS, JBSS, JOUT, KBH, KFRC, KLIC, KMX, LAD, LAUR, LCUT, LEN, LEN/B, LFCR, LGND, LNC, LNG, LNW, LPX, LXP, LZB, M, MAC, MBI, MCRI, MCS, MGM, MGPI, MHO, MLR, MOD, MPAA, MRTN, MSTR, MTG, MTH, MTW, MTZ, MVIS, NBIX, NBR, NC, NEU, NG, NHC, NKTR, NR, NRG, ODP, OHI, OI, OII, OPCH, OPY, OSBC, OSIS, PAG, PATK, PBI, PCG, PENN, PLAB, PLCE, PLUS, PPBI, PPC, PRDO, PTSI, PVH, PWR, RCII, RCKY, RCL, REX, RHP, RICK, RPT, RRC, RWT, SAH, SANM, SBGI, SCI, SCVL, SITC, SKYW, SLG, SLGN, SLM, SLP, SM, SMP, SPNS, SSP, STAA, STAR, STRS, STT, SVC, SWN, TCX, TEX, TGI, THC, THRM, TILE, TK, TPC, TRC, TTI, TUP, TWI, UFI, UFPT, UHAL, UHAL/B, UIS, UNFI, URI, VCEL, VERU, VHI, VSEC, VTR, VXRT, WDC, WINA, WMB, WNC, X, XRX, ZEUS \\
         B & \tiny AAL, ARCH, AREN, BLFS, CBL, CHK, CPE, CPSS, CRK, CTIC, DXLG, DXPE, HDSN, HOV, HSKA, LEU, LXU, MED, NOTV, NVAX, POWW, RAD, ROCC, SIRI, STRL, TCI, TRNS, VTNR, WFRD \\
    \end{tabular}
    \caption{Constituents of credit portfolios in dataset (ratings: Baa, Ba, B).}
    \label{tab:tickers2}
\end{table}

\section{Calibration of factor copulas}
\label{sec:fitting_copulas}

Classical parametric copulas - as used in actuarial science, risk aggregation in finance and insurance, biology, etc - model observed features, so that it's possible to calibrate them by likelihood maximization. In the case of credit factor copula models, instead, the copula is meant to model unobserved factors, so that the calibration is a more delicate issue.

Financial practitioners have developed several calibration techniques for factor copula models based on simplifying assumptions. The most common calibration technique applied to single-factor Gaussian and t copulas relies on the so-called large homogeneous portfolio (LHP) approximation. Under this approximation one assumes that the portfolio is sufficiently large to justify a Gaussian approximation for the losses distribution and that the obligors are homogeneous or exchangeable (i.e. correlations are assumed constant and identical for all pairs of obligors). It follows that the prices of many credit instruments admit a closed-form expression (see \cite{vasicek2002distribution} and \cite{schloegl2005note} for computations of CDO prices for single-factor Gaussian and t-copulas) so that the model can be calibrated under a risk-neutral measure directly to market prices. This calibration technique applies exclusively to exchangeable single-factor copulas and typically leads to very poor fits, because a single equicorrelation coefficient is unable to reproduce the observed market prices \cite{burtschell2005comparative}.

Other calibration techniques attempt to fit the equicorrelation coefficient directly to a target default correlation value, which is pre-specified by the practitioner (see \cite{bolder2018credit} for details). This procedure was common in the pricing and rating of newly issued CDOs in the early 2000s, but is ultimately arbitrary and not data-driven.

In this paper we propose a new procedure which can be applied to calibrate multi-factor Gaussian and t copula models directly on default probability data. 

Depending on the application at hand, these default probabilities can be either physical or risk-neutral probabilities. Physical default probabilities are normally used in a risk management or insurance context, while risk-neutral default probabilities are used in the pricing and hedging of credit derivatives. Our method can be applied in both cases.

We emphasize that our procedure does not rely on simplifying assumptions - such as the LHP approximation or the equicorrelation assumption - and readily extends to the case of multi-factor copula models. This methodology has been implemented in Python using automatic differentiation from the TensorFlow package and it has been validated on synthetically generated datasets. The code is freely available on GitHub\footnote{Link: \texttt{https://github.com/gvisen/credit-RBM-model}}.

\subsection{Multi-factor Gaussian copula calibration}
\label{subsec:gaussian_copula_calibration}

We briefly present the multi-factor Gaussian copula model, in order to fix our notation. In a multi-factor Gaussian copula model for a portfolio of $n$ debtors the unobserved (standardized) asset value of obligor $k$ at maturity is assumed to be of the following form:

$$ X_k = \sum_{i=1}^d a_{k,i} Z_i + \varepsilon_k \sqrt{1 - \sum_{i=1}^d a_{k,i}^2}, \quad k = 1, \ldots, n$$

where $Z = (Z_1, \ldots, Z_d)$ is a $d$-dimensional standard Gaussian vector, $\varepsilon = (\varepsilon_1, \ldots, \varepsilon_n)$ is an $n$-dimensional standard Gaussian vector of firm-dependent noise terms, and the vectors $Z$ and $\varepsilon$ are assumed to be independent.

The coefficients $(a_{k,i})$ are called factor loadings and implicitly encode the dependence of each obligor's default on the unobserved systematic factors $Z$.

The $k$-th obligor is assumed to default if its asset value at maturity is below a certain threshold $L_k$. The probability of default (PD) is therefore:

\begin{align*}
    \text{PD}_k & = \mathbb{P} \left( X_k < L_k \right) \\ 
    & = \mathbb{P} \left( \varepsilon_k < \frac{L_k - \sum_{i=1}^d a_{k,i} Z_i}{\sqrt{1 - \sum_{i=1}^d a_{k,i}^2}} \right) \\
    & = \Phi \left( \frac{L_k - \sum_{i=1}^d a_{k,i} Z_i}{\sqrt{1 - \sum_{i=1}^d a_{k,i}^2}} \right) \\
    & = \Phi \left( \frac{\Phi^{-1}(\bar{p}_k) - \sum_{i=1}^d a_{k,i} Z_i}{\sqrt{1 - \sum_{i=1}^d a_{k,i}^2}} \right)
\end{align*} 

where $\Phi$ is the cumulative distribution function of a standard Gaussian variable and where, in the last step, we have noticed that the value of the threshold $L_k$ is fully determined by the expected default probability of the obligor, which we denoted by $\bar{p}_k$. 

Suppose we are given a dataset of default probabilities, $(\text{PD}_1^{(\ell)}, \ldots, \text{PD}_n^{(\ell)})_{\ell = 1}^M$, where $M$ is the number of observations in our dataset (each observation $(\text{PD}_1^{(\ell)}, \ldots, \text{PD}_1^{(\ell)})$ is understood to be generated by an (unobserved) realization $Z^{(\ell)} = (Z_1^{(\ell)}, \ldots, Z_d^{(\ell)})$ of the vector of factors $Z$).

The average default probability $\bar{p}_k$ can be estimated directly from the sample mean on our dataset and is therefore unproblematic. In order to estimate the factor loadings, instead, we will require the following auxiliary variables:

$$ Y_k^{(\ell)} := \Phi^{-1}(\text{PD}_k^{(\ell)}) = \frac{\Phi^{-1}(\bar{p}_k) - \sum_{i=1}^d a_{k,i} Z_i^{(\ell)}}{\sqrt{1 - \sum_{i=1}^d a_{k,i}^2}}.$$

The vector $Y = (Y_1, \ldots, Y_n)$ is an affine transformation of the standard Gaussian vector $Z$ and is therefore a multivariate Gaussian vector itself, with mean $\mu = \Phi^{-1}(\bar{p}_k) / \sqrt{1 - \sum_{i=1}^d a_{k,i}^2}$ and covariance matrix $\Sigma = A A^T$, where we have defined $A_{k,i} := a_{k,i} / \sqrt{1 - \sum_{i=1}^d a_{k,i}^2}.$

The maximum likelihood estimator for the covariance matrix $\Sigma$ is the sample covariance matrix $\hat{\Sigma}$ of the vector $Y$, which can be computed from our dataset of default probabilities. In order to impose that $Y$ has the required factor structure, we must find a matrix $A$ such that $\hat{\Sigma} = A A^T$, which is a well-known problem in the field of statistical factor analysis. The problem is in general ill-posed, but can be solved, for instance, by minimizing the $L^2$ norm of the difference between the two matrices, yielding the following estimator:

$$ \hat{A} = \argmin_{A \in \mathbb{R}^{n \times d}} \| \hat{\Sigma} - AA^T \|_2^2.$$

The minimization can be performed numerically by gradient descent.

\subsection{Multi-factor t copula calibration}
\label{subsec:t_copula_calibration}

In a multi-factor t copula model the unobserved (standardized) asset values are assumed to be of the following form:

$$ X_k = \sqrt{\frac{\nu}{W}} \left( \sum_{i=1}^d a_{k,i} Z_i + \varepsilon_k \sqrt{1 - \sum_{i=1}^d a_{k,i}^2} \right), \quad k = 1, \ldots, n$$

where $Z$ and $\varepsilon$ are as in the Gaussian copula case, $W$ is an independent $\chi^2(\nu)$ random variable and $\nu > 0$ is a degrees of freedom parameter. This parametrization implies that the vector $X = (X_1, \ldots, X_n)$ is jointly t-distributed.

The probabilities of default (PDs) can be computed as before and are:

\begin{align*}
    \text{PD}_k & =\Phi \left( \frac{ \sqrt{\frac{W}{\nu}}F_\nu^{-1}(\bar{p}_k) - \sum_{i=1}^d a_{k,i} Z_i}{\sqrt{1 - \sum_{i=1}^d a_{k,i}^2}} \right)
\end{align*} 

where $F_\nu$ is the cumulative distribution function of a standard Student t variable with $\nu$ degrees of freedom. 

The calibration procedure is analogous to what we saw for the multi-factor Gaussian copula in Section \ref{subsec:gaussian_copula_calibration}. The auxiliary variables are:

$$ Y_k^{(\ell)} := \Phi^{-1}(\text{PD}_k^{(\ell)}) = \frac{\sqrt{\frac{W}{\nu}} F_\nu^{-1}(\bar{p}_k) - \sum_{i=1}^d a_{k,i} Z_i^{(\ell)}}{\sqrt{1 - \sum_{i=1}^d a_{k,i}^2}}.$$

For notational convenience, let us define $A_{k,i} := a_{k,i} / \sqrt{1 - \sum_{i=1}^d a_{k,i}^2}$ and $b_k := \sqrt{1 - \sum_{i=1}^d a_{k,i}^2} = \sqrt{1 / (1 + \sum_{i=1}^d A_{k,i})}$. Then the vector $Y = (Y_1, \ldots, Y_n)$ must satisfy the following moment conditions:

$$ \mathbb{E}[Y_k] = \frac{F_\nu^{-1}(\bar{p}_k)}{\sqrt{\nu} b_k} \mathbb{E}[\sqrt{W}] = \frac{F_\nu^{-1}(\bar{p}_k)}{b_k} \sqrt{\frac{2}{\nu}} \frac{\Gamma\left(\frac{\nu+1}{2}\right)}{\Gamma\left(\frac{\nu}{2}\right)}, \quad \forall k = 1, \ldots, n,$$

$$ \mathbb{E}[Y_k Y_j] = \frac{F_\nu^{-1}(\bar{p}_k) F_\nu^{-1}(\bar{p}_j)}{b_k b_j}  + \sum_{i=1}^d A_{k,i} A_{j,i}, \quad \forall k, j = 1, \ldots, n,$$

from which we obtain an analytical expression $\Sigma(A, \nu)$ for the covariance matrix of $Y$ as a function of $A$ and $\nu$. We can then calibrate these parameters by minimizing the $L^2$ norm of the difference between the empirical covariance matrix $\hat{\Sigma}$ of $Y$ and its analytical expression $\Sigma(A, \nu)$, yielding the following estimators:

$$ (\hat{A}, \hat{\nu}) = \argmin_{(A, \nu) \in \mathbb{R}^{n \times d} \times (0, +\infty)} \|\hat{\Sigma} - \Sigma(A, \nu)\|_2^2.$$

The minimization can be performed numerically by gradient descent.

\section{Results for all rating classes}
\label{subsec:results_all_rating_classes}

\subsection{Estimation of tail functions}
\label{subsec:tail_comparison_all_rating_classes}

We report below the results of Section \ref{sec:estimation_of_risk_measures}, specifically Figure \ref{fig:tail_comparison_copulas}, for all rating classes in our dataset.

\begin{figure}[H]
    \centering
    \includegraphics[width=\textwidth]{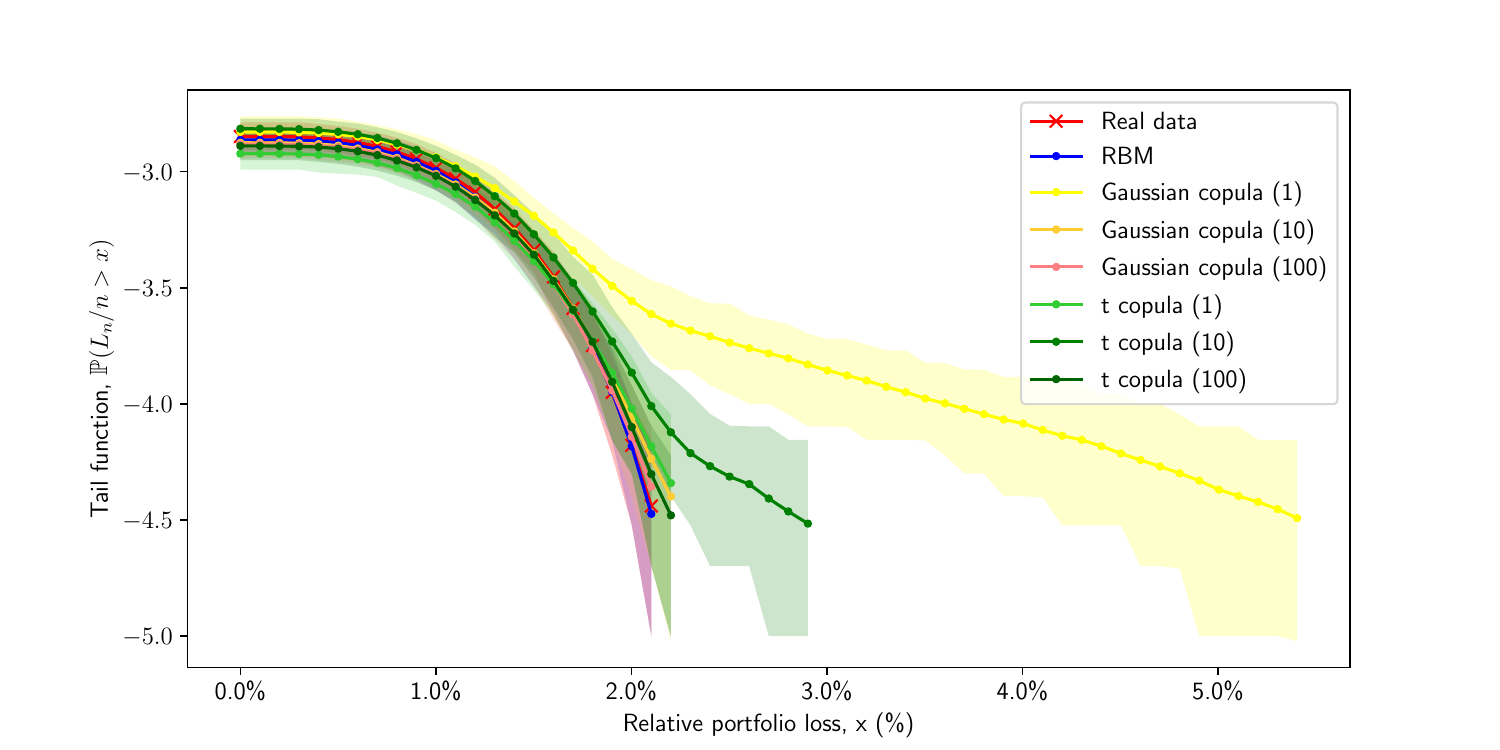}
    \caption{Rating Aaa.}
    \label{fig:tail_comparison_copulas_Aaa}
\end{figure}

\begin{figure}[H]
    \centering
    \includegraphics[width=\textwidth]{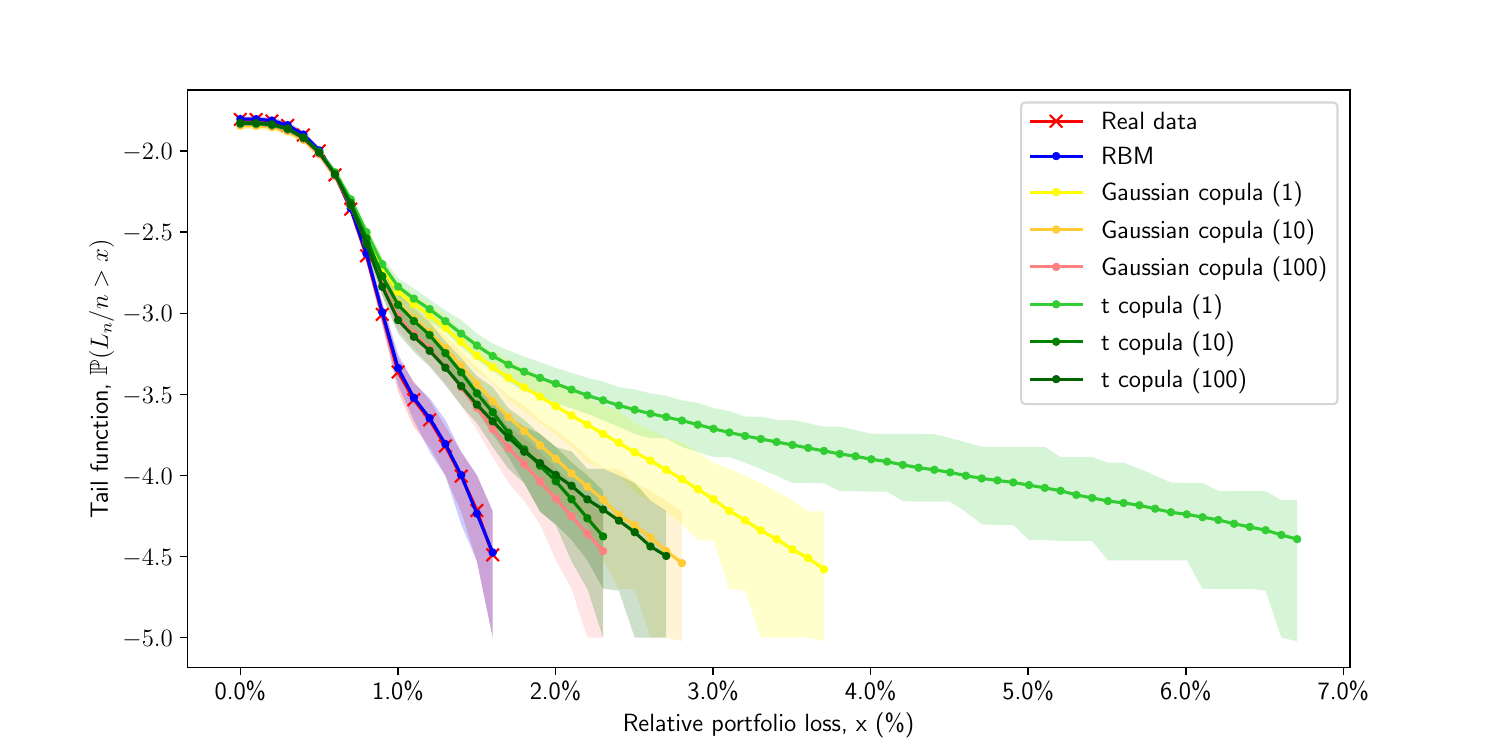}
    \caption{Rating Aa.}
    \label{fig:tail_comparison_copulas_Aa}
\end{figure}

\begin{figure}[H]
    \centering
    \includegraphics[width=\textwidth]{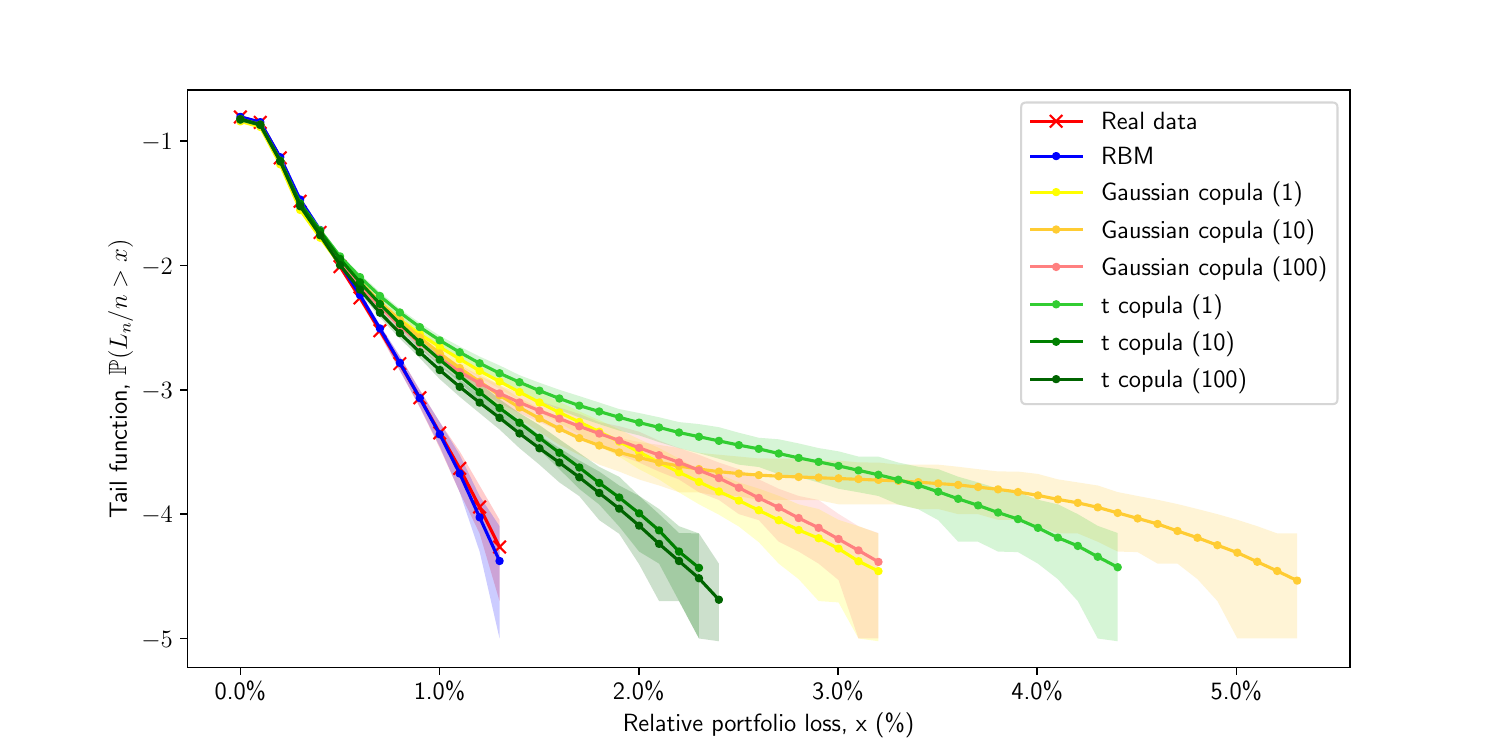}
    \caption{Rating A.}
    \label{fig:tail_comparison_copulas_A}
\end{figure}

\begin{figure}[H]
    \centering
    \includegraphics[width=\textwidth]{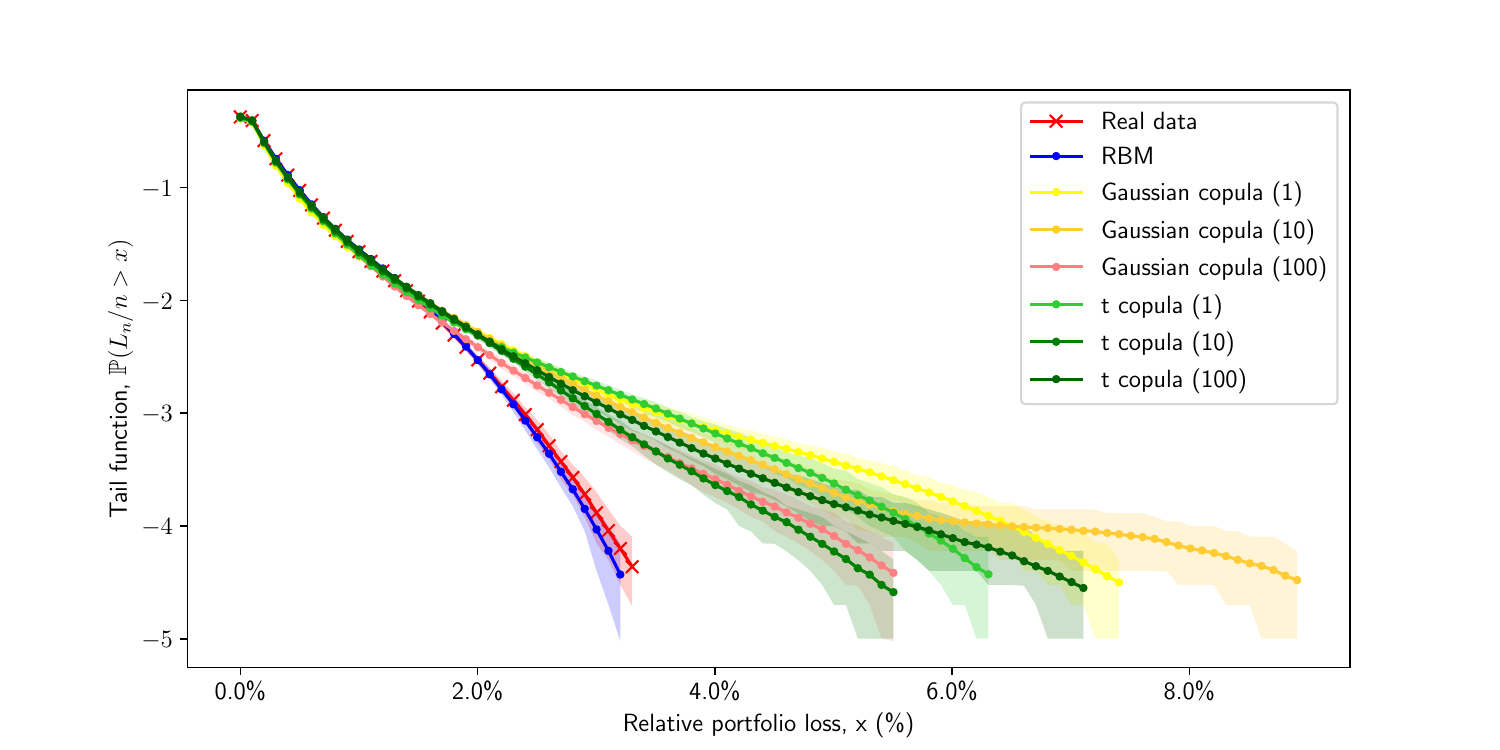}
    \caption{Rating Baa.}
    \label{fig:tail_comparison_copulas_Baa}
\end{figure}

\begin{figure}[H]
    \centering
    \includegraphics[width=\textwidth]{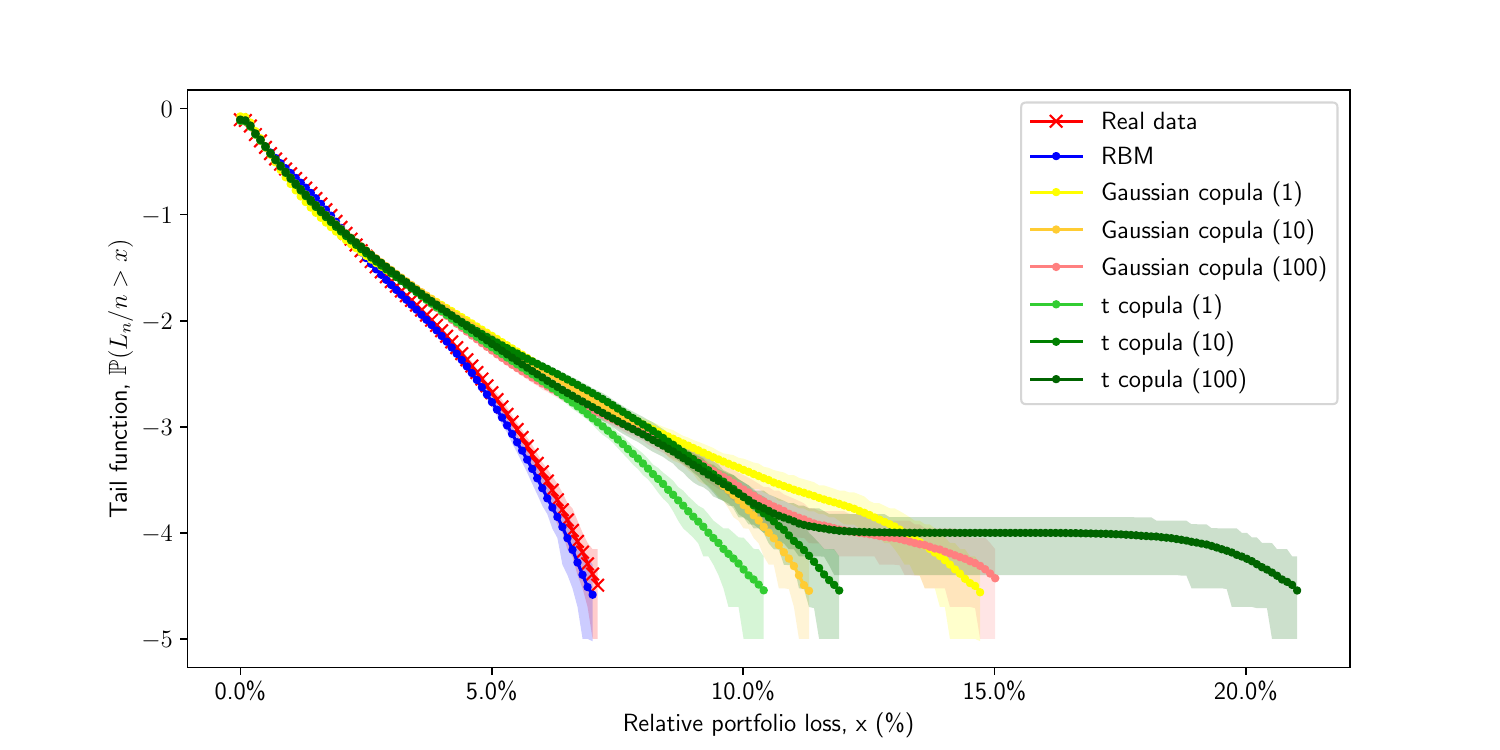}
    \caption{Rating Ba.}
    \label{fig:tail_comparison_copulas_Ba}
\end{figure}

\begin{figure}[H]
    \centering
    \includegraphics[width=\textwidth]{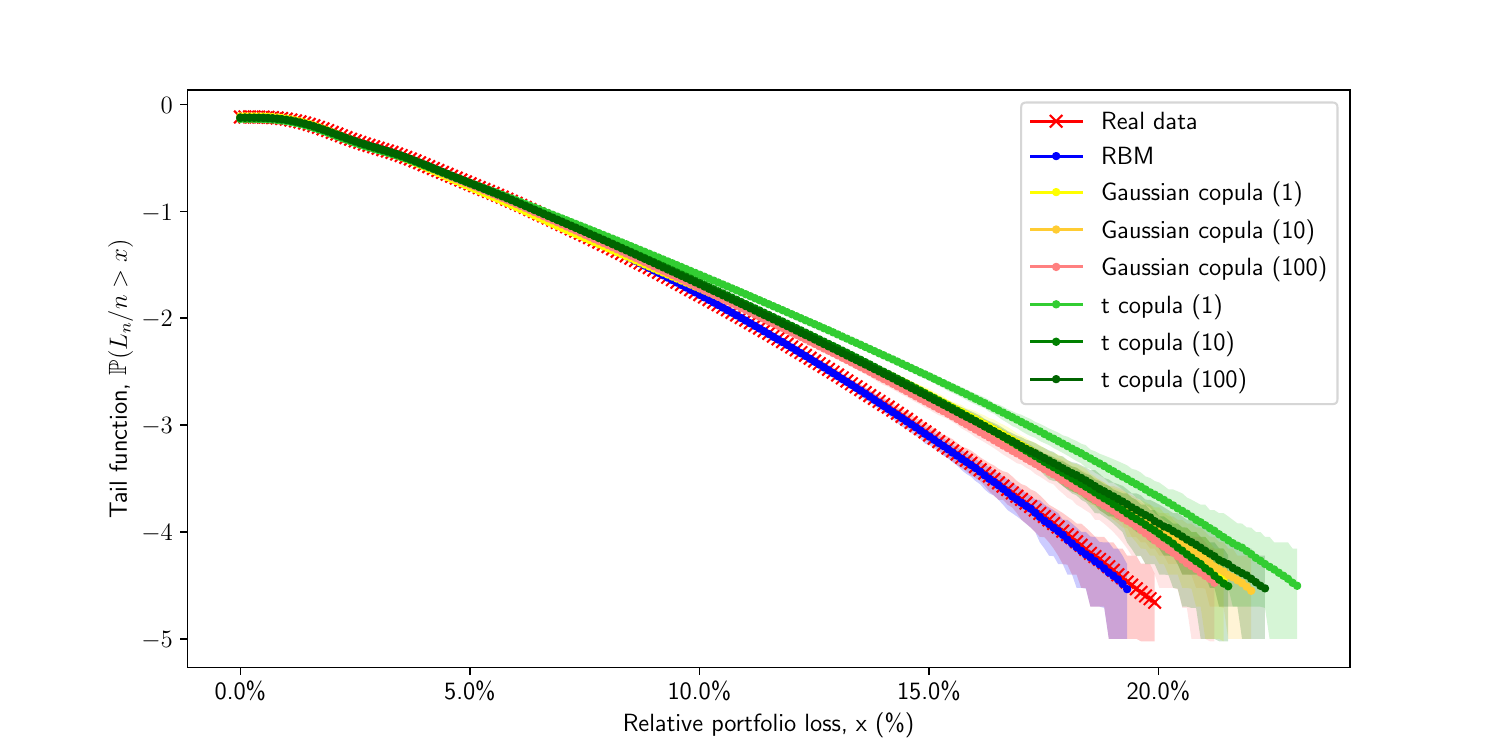}
    \caption{Rating B.}
    \label{fig:tail_comparison_copulas_B}
\end{figure}

\subsection{Stressed Value at Risk}
\label{subsec:stressed_var_all_rating_classes}

We report below the results of Section \ref{sec:stress_testing}, specifically Figure \ref{fig:stress_test_vars}, for all rating classes in our dataset.

\begin{figure}[H]
    \centering
    \includegraphics[width=\textwidth]{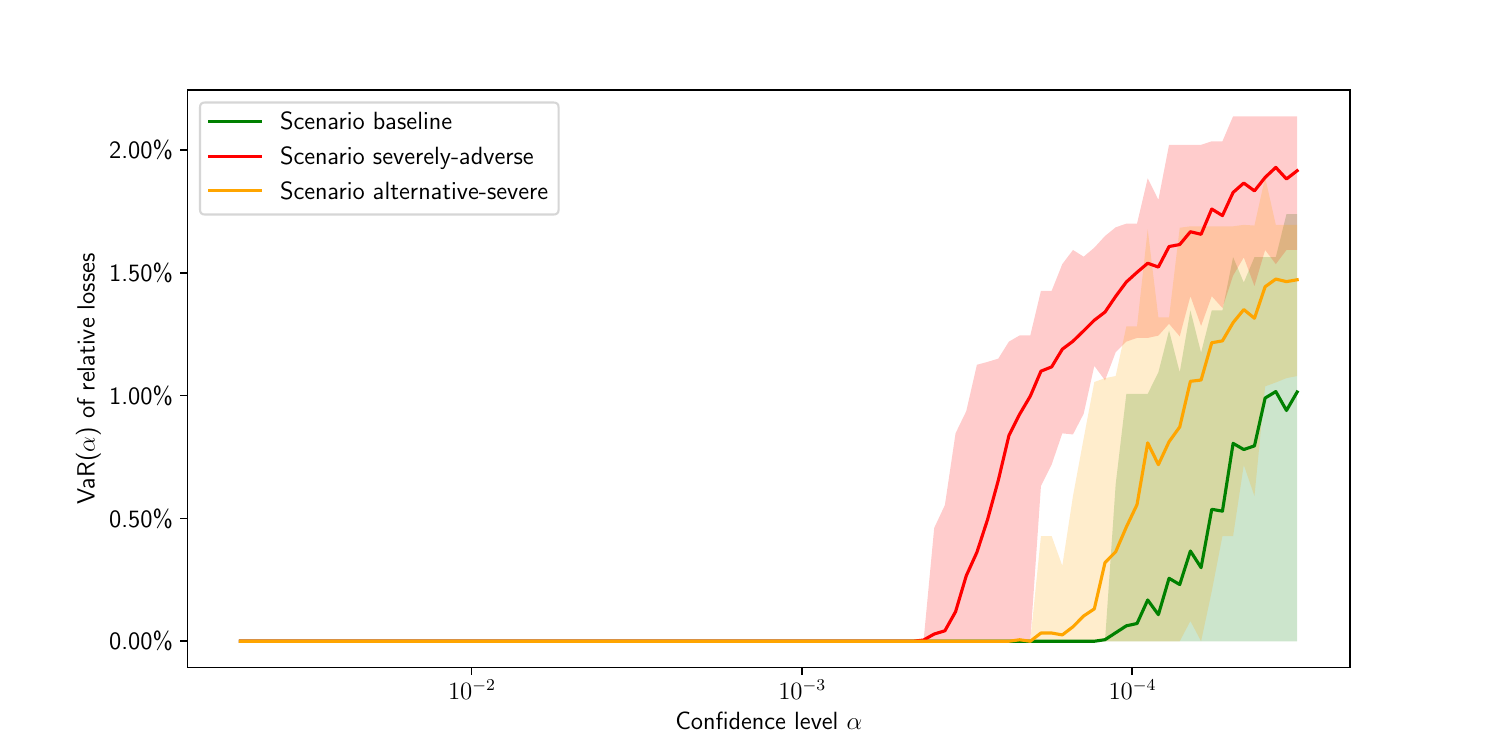}
    \caption{Rating Aaa.}
    \label{fig:stress_test_vars_Aaa}
\end{figure}

\begin{figure}[H]
    \centering
    \includegraphics[width=\textwidth]{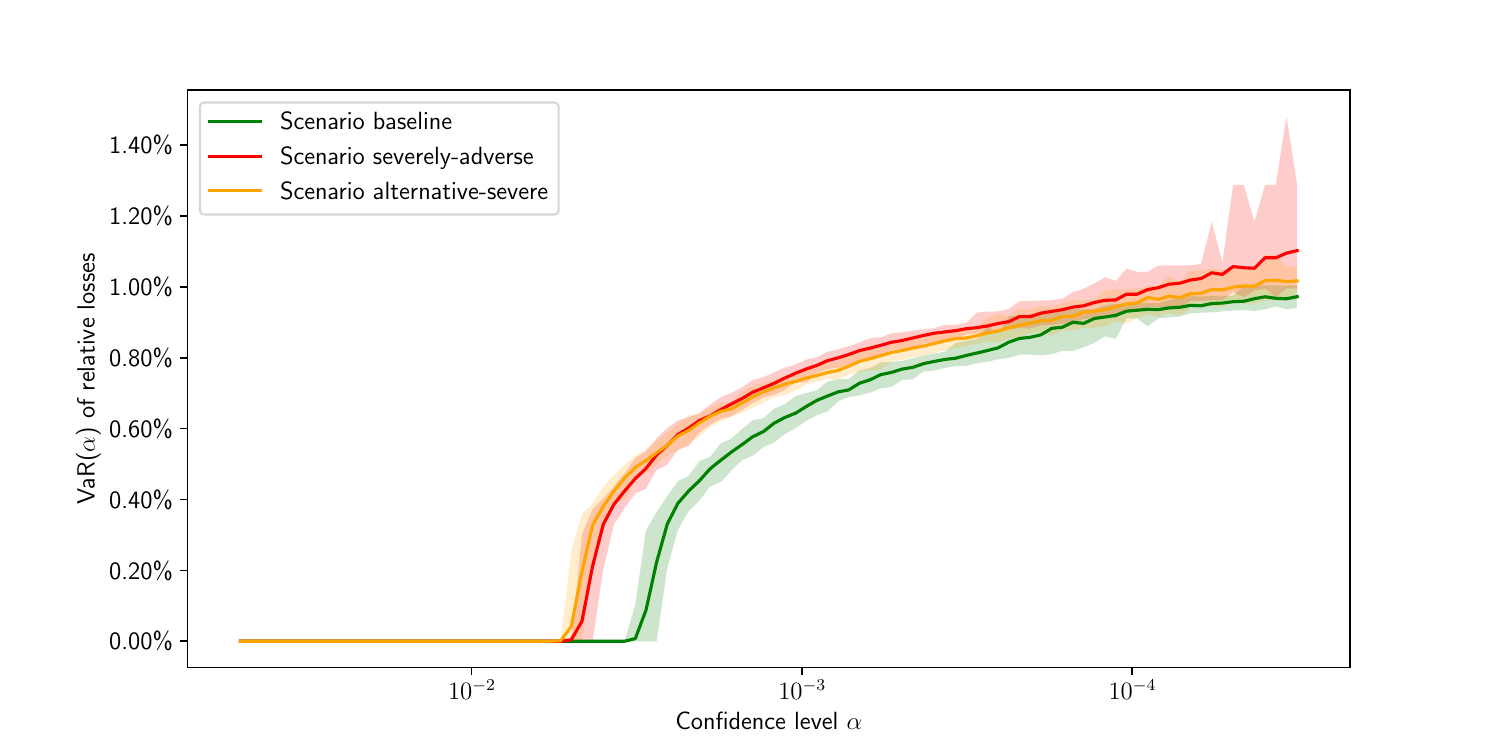}
    \caption{Rating Aa.}
    \label{fig:stress_test_vars_Aa}
\end{figure}

\begin{figure}[H]
    \centering
    \includegraphics[width=\textwidth]{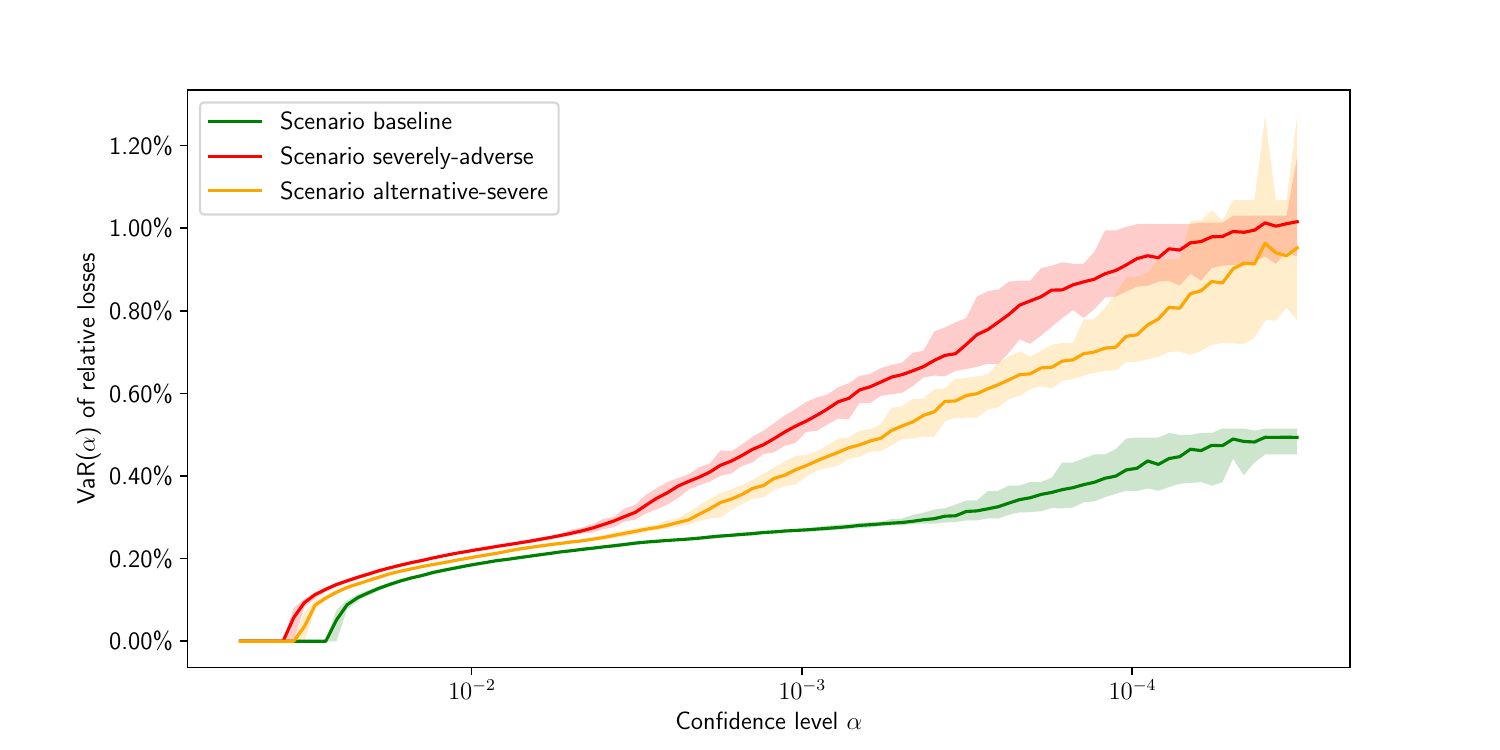}
    \caption{Rating A.}
    \label{fig:stress_test_vars_A}
\end{figure}

\begin{figure}[H]
    \centering
    \includegraphics[width=\textwidth]{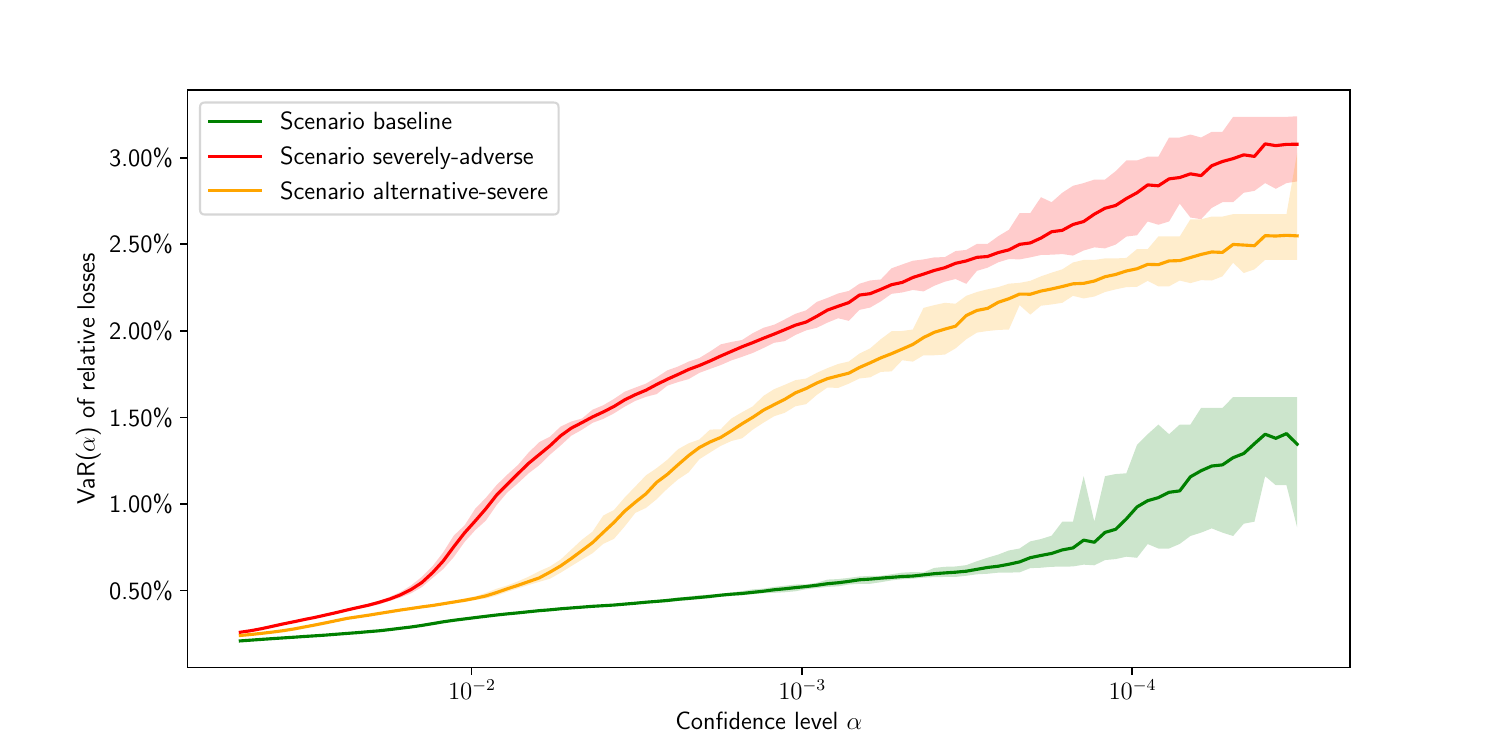}
    \caption{Rating Baa.}
    \label{fig:stress_test_vars_Baa}
\end{figure}

\begin{figure}[H]
    \centering
    \includegraphics[width=\textwidth]{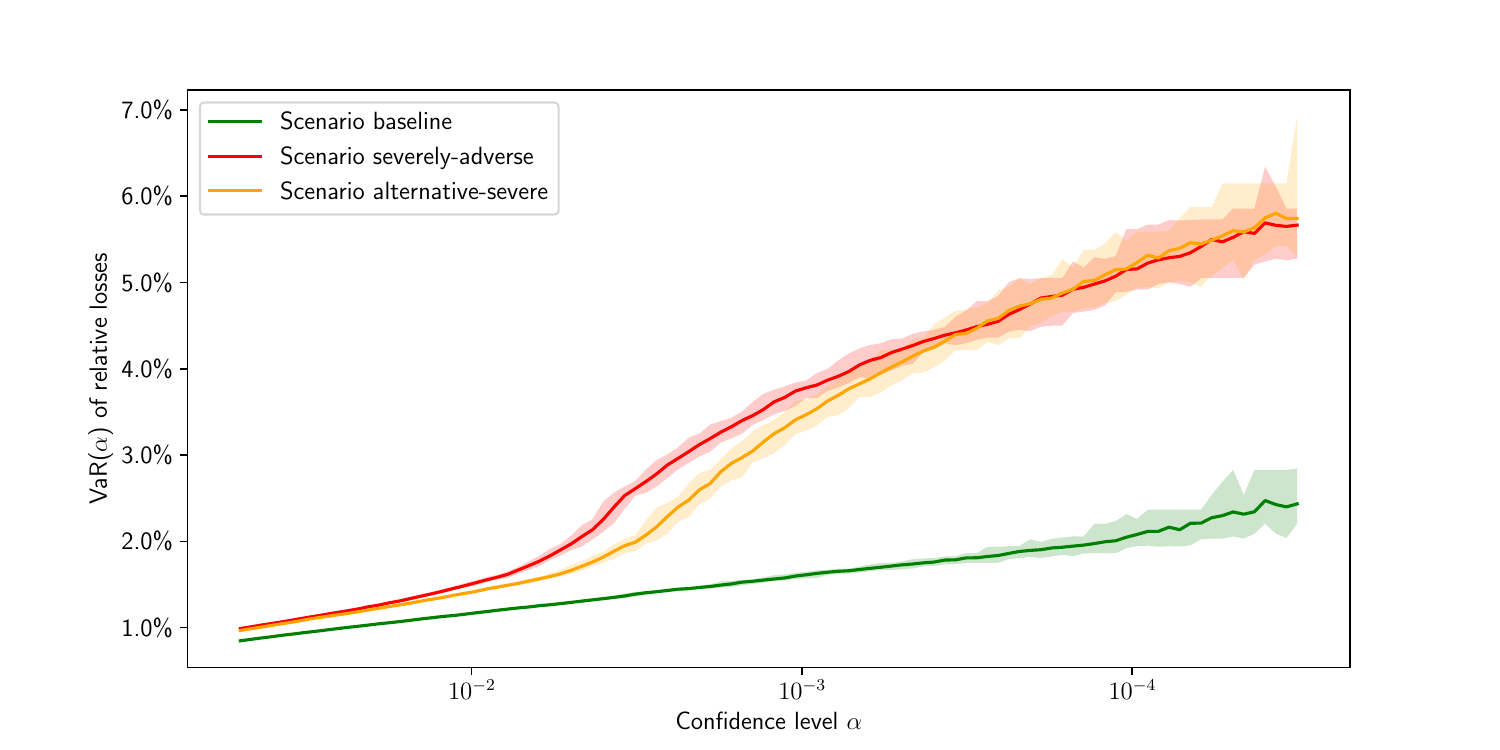}
    \caption{Rating Ba.}
    \label{fig:stress_test_vars_Ba}
\end{figure}

\begin{figure}[H]
    \centering
    \includegraphics[width=\textwidth]{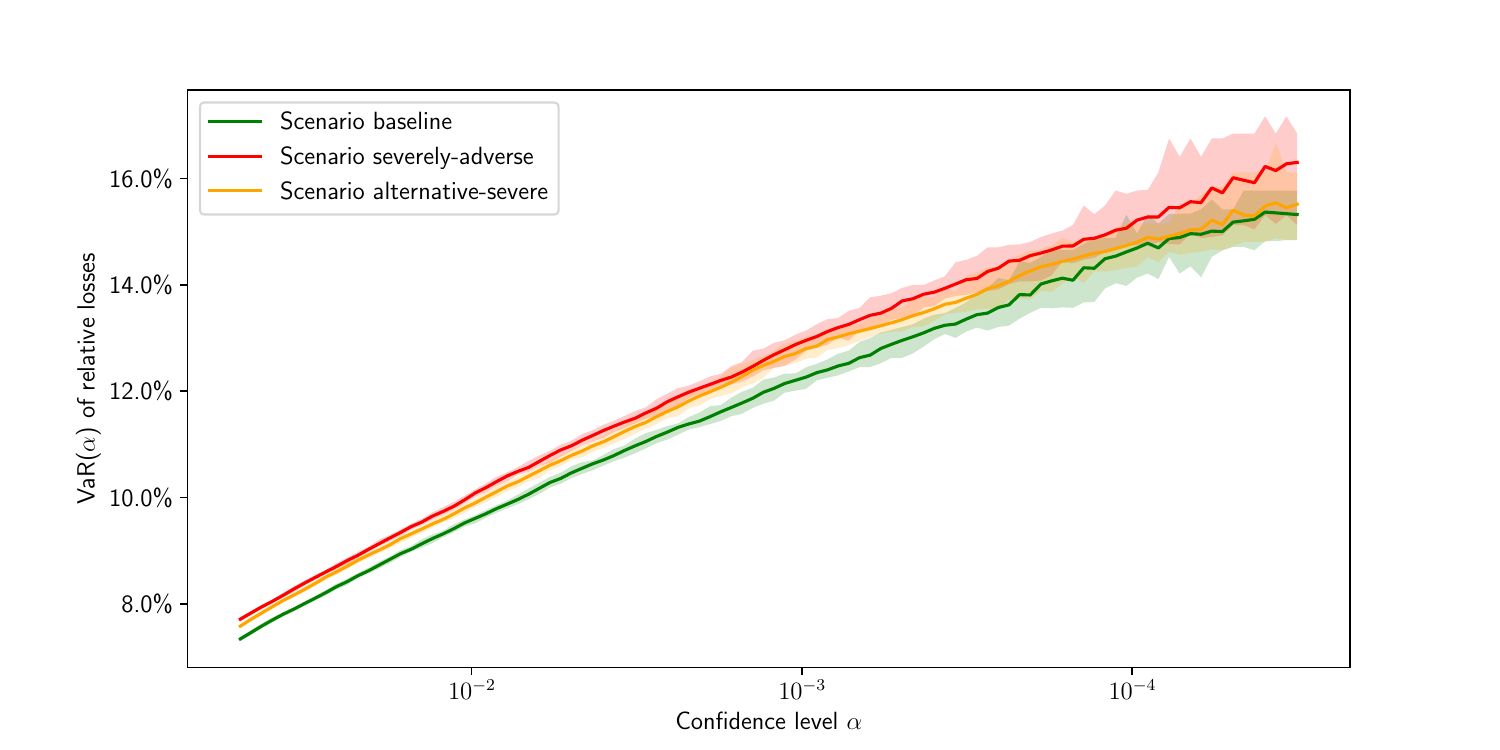}
    \caption{Rating B.}
    \label{fig:stress_test_vars_B}
\end{figure}

\bibliographystyle{amsplain}

\bibliography{refs.bib}

\end{document}